\documentclass[doublecolumn,american,english]{IEEEtran}
\usepackage[T1]{fontenc}
\usepackage{babel}
\usepackage{amsthm}
\usepackage{amsmath}
\usepackage{amssymb}
\usepackage{graphicx}
\usepackage[unicode=true,
 bookmarks=true,bookmarksnumbered=true,bookmarksopen=true,bookmarksopenlevel=1,
 breaklinks=false,pdfborder={0 0 0},backref=false,colorlinks=false]
 {hyperref}
\hypersetup{pdftitle={Your Title},
 pdfauthor={Your Name},
 pdfpagelayout=OneColumn, pdfnewwindow=true, pdfstartview=XYZ, plainpages=false}

\makeatletter
 \let\oldforeign@language\foreign@language
 \DeclareRobustCommand{\foreign@language}[1]{%
   \lowercase{\oldforeign@language{#1}}}
\theoremstyle{plain}
\newtheorem{thm}{\protect\theoremname}
\theoremstyle{remark}
\newtheorem{rem}[thm]{\protect\remarkname}
\theoremstyle{plain}
\newtheorem{lem}[thm]{\protect\lemmaname}
\theoremstyle{plain}
\newtheorem{cor}[thm]{\protect\corollaryname}
\theoremstyle{plain}
\newtheorem{prop}[thm]{\protect\propositionname}

\ifCLASSOPTIONcompsoc
\usepackage[caption=false,font=normalsize,labelfont=sf,textfont=sf]{subfig}
\else
\usepackage[caption=false,font=footnotesize]{subfig}
\fi

\newtheorem*{lemma*}{Lemma}

\makeatother

\addto\captionsamerican{\renewcommand{\corollaryname}{Corollary}}
\addto\captionsamerican{\renewcommand{\lemmaname}{Lemma}}
\addto\captionsamerican{\renewcommand{\propositionname}{Proposition}}
\addto\captionsamerican{\renewcommand{\remarkname}{Remark}}
\addto\captionsamerican{\renewcommand{\theoremname}{Theorem}}
\addto\captionsenglish{\renewcommand{\corollaryname}{Corollary}}
\addto\captionsenglish{\renewcommand{\lemmaname}{Lemma}}
\addto\captionsenglish{\renewcommand{\propositionname}{Proposition}}
\addto\captionsenglish{\renewcommand{\remarkname}{Remark}}
\addto\captionsenglish{\renewcommand{\theoremname}{Theorem}}
\providecommand{\corollaryname}{Corollary}
\providecommand{\lemmaname}{Lemma}
\providecommand{\propositionname}{Proposition}
\providecommand{\remarkname}{Remark}
\providecommand{\theoremname}{Theorem}

\begin{document}

\title{The Bethe Free Energy Allows to Compute the Conditional Entropy of
Graphical Code Instances. A Proof from the Polymer Expansion}


\author{Nicolas~Macris and~Marc~Vuffray%
\thanks{Nicolas~Macris is with the School of Computer and Communication Science,
Ecole Polytechnique Fédérale de Lausanne, Lausanne, Switzerland, e-mail:
\protect\href{mailto:nicolas.macris@epfl.ch}{nicolas.macris@epfl.ch}.%
}%
\thanks{Marc~Vuffray is with the Theory Division and Center for Nonlinear
Studies, Los Alamos National Laboratory, Los Alamos NM, USA, e-mail:
\protect\href{mailto:vuffray@lanl.gov}{vuffray@lanl.gov}.%
}}

\maketitle
\begin{abstract}
The main objective of this paper is to explore the precise relationship 
between the Bethe free energy (or entropy) and the Shannon conditional 
entropy of graphical error
correcting codes. 
The main result shows that the Bethe free energy
associated with a low-density parity-check code used over a binary
symmetric channel in a large noise regime is, with high probability,
asymptotically exact as the block length grows. To arrive at this result we 
develop new techniques for rather general graphical models based on the loop sum as a starting point and 
 the polymer
expansion from statistical mechanics.
The true free energy is computed as a series expansion containing
the Bethe free energy as its zero-th order term plus
a series of corrections. It is easily seen that convergence criteria
for such expansions are satisfied for general high-temperature models.
We apply these general results
to ensembles of low-density generator-matrix and parity-check codes.
While the application to generator-matrix codes follows standard ``high temperature''
methods, the case of parity-check codes requires non-trivial new ideas
because the hard constraints correspond to a zero-temperature regime.
Nevertheless one can combine the polymer expansion with expander and
counting arguments to show that the difference between the true and
Bethe free energies vanishes with high probability in the large block
length limit. \end{abstract}


\IEEEpeerreviewmaketitle{}

\section{Introduction\label{sec:introduction}}

\IEEEPARstart{O}{ften} one needs to compute the free energy and/or
entropy of a random graphical model. For example in the theory of
codes on graphs, which is our main motivation here, it is known that
the conditional input-output Shannon entropy of a graphical code used
over a binary memoryless symmetric channel is related by a simple
formula to the free energy of the graphical model arising in Maximum
Posterior decoding. The Bethe approximation and the related Belief
Propagation (BP) equations may sometimes offer a good starting point for
computing this free energy. However it is seldom a controlled approximation.
In special cases, it can be a rigorous upper or lower bound but neither
holds in general. A very interesting general result of Vontobel \cite{vontobel2010counting}
relates the Bethe free energy of an instance of a graphical model
to the average of the true free energy over all graph covers of the
instance. In the special case of Ising-like graphical models with
attractive pair interactions Wainwright \cite{sudderth2007loop} has
shown that, under additional special conditions, the Bethe free energy
is a bound to the true free energy. This has been extended recently
to a much wider setting (for interactions satisfying log-supermodularity
conditions) by Ruozzi \cite{ruozzi2012bethe}. For counting independent
sets in sparse graphs with large girth, Chandrasekaran et al. \cite{chandrasekaran2011counting}
show that the Bethe free energy is asymptotically exact as the size
of the graph grows. References \cite{sudderth2007loop} and \cite{chandrasekaran2011counting}
use a generic representation of the partition function developed by
Chertkov and Chernyak \cite{chertkov2006loop} and which also forms
the basis of this work.

In this paper our main objective is to show that the Bethe free energy
associated with a low-density parity-check (LDPC) code used over a
binary symmetric channel (BSC) in a large-noise regime is, with high
probability, asymptotically exact as the block length grows. In that regime 
the Bethe free energy allows to compute the Shannon entropy of the input
code word conditioned on the output message.
Admittedly the high noise 
regime is not the most interesting one for practical purposes; however it is the regime 
where the conditional Shannon entropy is non-trivial; indeed for low noise it vanishes
in the large block length limit. It is conceptually interesting that the Bethe free energy
and the Shannon conditional entropy are intimately related in the regime where they are non trivial. 
Our proofs work for high enough noise levels but presumably the result holds  
for all noise levels above the MAP threshold of the code ensemble. 

We introduce techniques that are somewhat new in the theory of graphical codes. 
The proof
is based on a tool from statistical mechanics, called the polymer
expansion (see the end of this introduction for related ideas). Interestingly
the polymer expansion has to be combined with special features of
the graphical model associated with LDPC codes (features that are
not needed in the usual applications of the polymer expansion). In
fact the polymer expansion has an easy application to the case of
low-density generator-matrix (LDGM) codes for high noise and more
generally to graphical models in a high-temperature regime. Since
we believe these tools are somewhat new to the coding theory community
we present these applications as well. This also serves the pedagogical
purpose of introducing polymer expansions.

Let us immediately mention that we develop the analysis for the BSC
only to keep the technicalities to a minimal level, but the present
techniques have a wider range of validity.

A few years ago Chertkov and Chernyak \cite{chertkov2006loop} developed
a loop-sum representation for the partition function of graphical
models. The virtue of this representation is that the partition function
factorizes as the product of the Bethe contribution and a finite sum
of terms over subgraphs (not necessarily connected) with no dangling
edges. Each term of the sum involves only belief propagation messages
adjacent to the subgraphs. In \cite{chertkov2006loop} these subgraphs
are called loops.

It is tempting to use the loop-sum representation not only as a mere
formal tool, but to compare the true and Bethe free energies. One
of the aims of this contribution is to develop this idea systematically.
We recognize that the loop sum is itself the partition function of
a system of polymers. A loop is the union of connected subgraphs with
no dangling edges, which are called polymers. Each polymer has an
associated weight which depends only on belief propagation messages
adjacent to it. By definition the polymers cannot intersect. This
places a constraint that can be viewed as an infinitely repulsive
pair interaction. The representation of the loop sum as the partition
function of a polymer system with infinitely repulsive interactions
opens the way to the computation of the logarithm of this sum via
a combinatorial expansion known in statistical mechanics as the polymer
expansion \cite{brydges1984short}. If this expansion converges, then
we can in principle, compute corrections to the Bethe free energy
(which corresponds to the zero-th order term of the expansion) to
an arbitrarily high order. If the girth of the graph is large all
contributions beyond the zero-th order Bethe free energy only come
from large loops and, if these contributions become small as the size
of a loop increases, one may expect that, provided the expansion converges
uniformly in system size, the Bethe free energy is asymptotically
exact. More generally this mechanism may occur for typical instances
of graphs from a random ensemble of Erd\H{o}s-Rényi type, because
the neighborhood of a given vertex is tree like. Conversely, when
the Bethe free energy is asymptotically exact one may hope that the
expansion converges and is controllable. This is of course not necessarily
true as cancellations between terms in the expansion may occur. On
the other hand we know of systems, such as random constraint satisfaction
models (e.g, $K$-SAT or $Q$-coloring) or spin glasses, where the
true free energy is definitely not given by the Bethe formula (even
when averaged over the graph ensemble). For these systems it is certainly
not possible for the polymer expansion to converge. The local tree-like
nature of the graph is not sufficient to eliminate the contributions
of large loops when long ranged correlations are present. 

The program outlined above is first carried out in various cases. While our main application is for LDPC codes, 
we also consider for pedagogical reasons high temperature models that 
have an immediate application to LDGM codes (at high noise).

For high temperature models the polymer expansion starts
with a zero-th order term and the rest of the series is absolutely
convergent provided the temperature is large enough. We show that
this has an application to models whose factor graph has
a large girth in the sense that the girth grows logarithmically with
the size of the graph. For such models the Bethe free energy is asymptotically
exact in the thermodynamic limit. Another application is to irregular
LDGM codes for large noise (no assumption on the girth). We show that the free energy of an instance
drawn at random from an irregular LDGM ensemble is, with high probability,
given by the Bethe formula in the large block length limit.

Let us now describe the results concerning LDPC codes. We consider
regular LDPC codes used over a BSC (no assumption on the girth). Our analysis goes through essentially
unchanged for irregular codes but we refrain to present it in such
generality to avoid technical complications. In the case of LDPC codes
we cannot prove that the polymer expansion is absolutely convergent.
The reason is that the check node constraints are not of high-temperature
nature but rather low (even zero) temperature. It is therefore not
clear a priori why the polymer expansion should be of any use, except
for the fact that the zero-th order term is the Bethe free energy.
However, interestingly, using expander properties of typical instances
from the LDPC ensemble we can show that a truncated form of this expansion
does converge absolutely (uniformly in the system size). Moreover
the convergent truncated expansion accounts for the biggest part of
the corrections to the Bethe free energy, up to a remainder of order
$O(e^{-n\epsilon})$, $\epsilon>0$. This remainder part is not expanded
but estimated by a combinatorial counting method. The final result
is again that the Bethe free energy is asymptotically exact with high
probability in the large size limit. 

Let us briefly comment on the connections of this work with other
recent approaches. For the class of graphical models that describe
communication with low-density parity-check and low-density generator-matrix
codes over binary-symmetric memoryless channels we have plenty of
evidence that the replica-symmetric solution\footnote{Replica-symmetric formulas are averaged forms of the Bethe formulas,
where the average is over the channel output realizations and code
ensemble.} is exact. Bounds between the replica-symmetric and true free energy
were derived in \cite{montanari2005tight}, \cite{macris2007griffith},
\cite{kudekar2009sharp}, and for the special case of the binary erasure
channel equality was proven in \cite{richardson2008modern}, \cite{korada2007exact}.
These results are based on specific methods such as combinatorial
calculations for the binary erasure channel, and the interpolation
method for the bounds on general channels. In \cite{kudekar2011decay}
a more generic approach is taken based on cluster expansions combined
with duality. The cluster expansions used in \cite{kudekar2011decay}
are sophisticated forms of polymer expansions. It is proven that correlations
between pairs of distant (with respect to graph distance) bits decay
exponentially fast for LDGM codes in the regime of large noise, and
LDPC codes in the regime of small noise. This also allowed to conclude
that the replica symmetric formulas are exact in these regimes for
general binary-symmetric memoryless channels. A case where the cluster
expansions in \cite{kudekar2011decay} do not work is that of LDPC
codes on general channels in the regime of large noise considered
here. We also stress that while in the case of LDGM codes at high noise one could also 
make use of the Dobrushin uniqueness condition (see e.g. \cite{Simon}) to prove correlation decay (and subsequently exactness
of the Bethe free energy), for LDPC codes this method breaks down. Indeed Dobrushin's condition 
fails in the presence of parity check hard constraints (see \cite{kudekar2011decay} for more details).
A preliminary account of the results and methods presented here was given in \cite{NMMVisit2012}.

In the next Section \ref{sec:preliminaries} we give the precise definitions
of the models and briefly review the associated Bethe formulas. The
main results pertaining to LDGM and LDPC codes are summarized in Section
\ref{sec:main results}. The polymer representation and expansion
are developed in Section \ref{sec:loop sum and polymer expansion}.
This expansion is then applied to the analysis of general factor graph
models and LDGM codes for large noise in Sections \ref{sec:high temperature models}
and \ref{sec:analysis for ldgm codes}. The more involved analysis
for LDPC codes is then presented in Section \ref{sec: analysis for ldpc codes}.
Extensions of the method presented in the paper are discussed in \ref{sec:discussion}.
For the convenience of the reader, simple derivations of the loop-sum
identity and polymer expansion are reviewed in a streamlined fashion
in appendices \ref{app:loop sum identity} and \ref{app:polymer}.
Other appendixes contain more technical material needed throughout
the analysis.

\section{Preliminaries\label{sec:preliminaries}}

\subsection{Factor Graphs\label{subsec:factor graphs}}

We begin with a few definitions and notations. Consider two vertex
sets: $V$ a set of $n$ variable nodes and $C$ a set of $m$ check
nodes. We think of $n$ and $m$ large. We consider bipartite graphs
- call them $\Gamma$ - connecting $V$ and $C$. The set of edges
is $E$. When we say that $\Gamma$ is random we mean that we draw
it uniformly from some specified ensemble. The corresponding expectation
and probability are denoted by $\mathbb{E}_{\Gamma}$, $\mathbb{P}_{\Gamma}$.
Letters $i,j$ will always denote nodes in $V$ and letters $a,b$
nodes in $C$. We reserve the notations $\partial i$ (resp. $\partial a$)
for the sets of nodes that are neighbors of $i$ (resp. $a$) in $\Gamma$.

For a graph $\Gamma$ from a standard ensemble LDGM$(\Lambda,P)$
\cite{richardson2008modern} the fraction of variable nodes of degree
$1\leq s\leq l_{{\rm max}}$ is $\Lambda_{s}\geq0$, and the fraction
of check nodes with degree $1\leq t\leq r_{{\rm max}}$ is $P_{t}\geq0$.
Of course $\sum_{s=1}^{l_{{\rm max}}}\Lambda_{s}=\sum_{t=1}^{r_{{\rm max}}}P_{t}=1$.
Here $\Gamma$ is the Tanner graph of an LDGM code with design rate
$r/l=n/m$, where $l$ and $r$ are the average variable and check
nodes degree respectively. The large block length limit corresponds
to $n,m\to\infty$ with fixed design rate.

For LDPC codes, we will limit ourselves to regular codes. Instead
of working with the standard LDPC$(l,r)$ ensemble with variable node
degree $l$ and check node degree $r$, we find it more convenient
to consider a different ensemble $\mathcal{B}(l,r,n)$. This is simply
the set of all bipartite $(l,r)$ regular graphs - call them $\Gamma$
- connecting $V$ and $C$. In other words vertices of $V$ have degree
$l$, vertices of $C$ have degree $r$, and there are no double edges.
$\Gamma\subset\mathcal{B}(l,r,n)$ is the Tanner graph of an LDPC
code with design rate $1-l/r=1-m/n$. The large block length limit
again corresponds to $n,m\to+\infty$ with fixed design rate.

In the case of LDPC codes we will make use of expansion arguments.
For the convenience of the reader we briefly review the necessary
tools \cite{richardson2008modern}. We will say that $\Gamma$ is
a $(\lambda,\kappa)$ expander if for every subset $\mathcal{V}\subset V$
such that $\left\vert \mathcal{V}\right\vert <\lambda n$ we have
\begin{equation}
\left\vert \partial\mathcal{V}\right\vert \geq\kappa l\left\vert \mathcal{V}\right\vert ,
\end{equation}
where $\vert\partial\mathcal{V}\vert$ is the number of check nodes
that are connected to $\mathcal{V}$, and $\lambda$, $\kappa$ are
two positive numerical constants. Take a random $\Gamma\subset\mathcal{B}\left(l,r,n\right)$.
Fix $0<\kappa<1-\frac{1}{l}$ and $0<\lambda<\lambda_{0}$ where $\lambda_{0}$
is the (only) positive solution of the equation\footnote{Here $h_{2}(x)=-x\ln x-(1-x)\ln(1-x)$ is the binary entropy function.}
\begin{equation}
\frac{l-1}{l}h_{2}(\lambda_{0})-\frac{1}{r}h_{2}(\lambda_{0}\kappa r)-\lambda_{0}\kappa rh_{2}(\frac{1}{\kappa r})=0\,.
\end{equation}
Then we have 
\begin{equation}
\mathbb{P}[\Gamma{\rm ~is~a~(\lambda,\kappa)~expander}]=1-O\bigl(\frac{1}{n^{l(1-\kappa)-1}}\bigr).
\end{equation}
Later on we need to take $\kappa\in]1-\frac{2(r-1)}{lr},1-\frac{1}{l}[$,
which is always possible for $r>2$. In the rest of the paper $\kappa$
is always a constant in this interval, and $0<\lambda<\lambda_{0}$.
For concreteness, one can think of the example $\left(l,r\right)=\left(3,6\right)$,
$\kappa=0.5$ and $\lambda_{0}=7.7\times10^{-4}$.

\subsection{General Factor Graph Models}

The LDGM and LDPC codes are special cases of general factor graph
models. We find it convenient to develop the formalism of the loop
sum and polymer expansions in a unified manner which applies to general
models.

Consider a bipartite graph $\Gamma$. We construct a general factor
graph model or spin system as follows. We attach spin degrees of freedom
$s_{i}\in\{-1,+1\}$ to nodes $i\in V$. A spin configuration is an
assignment $\underline{s}=(s_{i})_{i\in V}$. To each check node $a$
we associate a weight depending on spins $i\in\partial a$. The collection
of spins $s_{i}$ with $i\in\partial a$ and the weight are denoted
$s_{\partial a}$ and $\psi_{a}(s_{\partial a})$. The partition function
of the factor graph model (or spin system) is 
\begin{equation}
Z=\sum_{\underline{s}\in\{-1,+1\}^{n}}\prod_{a\in C}\psi_{a}(s_{\partial a}).\label{eq:spinsystem}
\end{equation}
The free energy is defined by 
\begin{equation}
f=\frac{1}{n}\ln Z
\end{equation}
and the thermodynamic limit is the limit $n\to+\infty$.

If we restrict ourselves to the class of strictly positive weights
their most general form is 
\begin{equation}
\psi_{a}(s_{\partial a})=\exp\{\beta\sum_{I\subset\partial a}J_{I}\prod_{i\in I}s_{i}\},
\end{equation}
where $\beta>0$ has the interpretation of an inverse temperature
and $J_{I}\in\mathbb{R}$ have the interpretation of coupling constants\footnote{Units are suitably chosen so that $\beta J_{I}$ is dimensionless.}.
When we speak of a high-temperature regime it is meant that $\beta>0$
is small enough so that 
\begin{equation}
\mu\equiv2\beta\sup_{a\in C}\sum_{I\subset\partial a}\vert J_{I}\vert<<1.
\end{equation}
We remark for later use that in a high-temperature regime 
\begin{equation}
\bigr\vert\psi_{a}(\{s_{\partial a}\})-1\bigl\vert\leq2\beta\sup_{a\in C}\sum_{I\subset\partial a}\vert J_{I}\vert=\mu.\label{eq:hightemperature}
\end{equation}
It will become clear that for LDGM codes the high-temperature regime
is equivalent to large noise. However for LDPC codes this is not true
because these codes essentially correspond to take $J_{I}=+\infty$.

\subsection{Transmission with LDGM Codes.}

We transmit codewords from an LDGM code with Tanner graph $\Gamma$
and uniform prior over a BSC with flip probability $p$. Here information
bits $\underline{u}=(u_{i})_{i=1}^{n}$ are attached to variable nodes
$V$ and codewords are given by $\underline{x}=(x_{a})_{a=1}^{m}$
with 
\begin{equation}
x_{a}=\oplus_{i\in\partial a}u_{i}\,.
\end{equation}
We must have $n<m$ and $l>r$ so that the design rate $r/l$ is well
defined. We can assume without loss of generality that the all-zero
codeword is transmitted. The posterior probability that $\underline{x}=(x_{i})_{i=1}^{n}\in\{0,1\}^{n}$,
or equivalently $\underline{u}=(u_{a})_{a=1}^{m}$, is transmitted
given that $\underline{y}=(y_{a})_{i=1}^{n}\in\{0,1\}^{n}$ is received,
reads 
\begin{equation}
p_{\underline{U}|\underline{Y}}\left(\underline{u}|\underline{y}\right)=\frac{1}{Z_{{\rm LDGM}}}\prod_{a\in C}e^{h_{a}\prod_{i\in\partial a}(-1)^{u_{i}}}.
\end{equation}
In this expression 
\begin{equation}
h_{a}=(-1)^{y_{a}}\frac{1}{2}\ln\frac{1-p}{p}
\end{equation}
are the half-log-likelihood variables and 
\begin{equation}
Z_{{\rm LDGM}}=\sum_{\underline{u}\in\{0,1\}^{n}}\prod_{a\in C}e^{h_{a}\prod_{i\in\partial a}(-1)^{u_{i}}}\label{eq:partitionLDGM}
\end{equation}
is the partition function. The amplitude of $h_{a}$ is set to 
\begin{equation}
\vert h_{a}\vert\equiv h\equiv\frac{1}{2}\ln\frac{1-p}{p}.
\end{equation}
It is good to keep in mind that the high-noise regime - $p$ close
to $1/2$ - corresponds to small $h$. It is equivalent to describe
the channel outputs $\underline{y}$ in terms of the half-log-likelihood
variables $\underline{h}=(h_{a})_{a=1}^{m}$ which are i.i.d with
probability distribution 
\begin{equation}
c(h_{a})=(1-p)\delta(h_{a}-h)+p\delta(h_{a}+h).
\end{equation}
The expectation with respect to this distribution is called $\mathbb{E}_{\underline{h}}$. 
\begin{rem}
Equ. \eqref{eq:partitionLDGM} is the partition function of a spin
system with one coupling constant $\beta J_{I}\to h_{a}$ per check,
and the high-temperature regime \eqref{eq:hightemperature} simply
corresponds to $h<<1$. 
\end{rem}
The free energy for fixed $(\Gamma,\underline{y})$ is 
\begin{equation}
f_{{\rm LDGM}}=\frac{1}{n}\ln Z_{{\rm LDGM}}
\end{equation}
For communications, the importance of this quantity stems from the
fact that it is intimately related to the Shannon conditional entropy
by the simple formula, 
\begin{equation}
\frac{1}{n}H_{\mathrm{LDGM}}\left(\underline{U}|\underline{Y}\right)=\mathbb{E}_{\underline{h}}\left[f_{{\rm LDGM}}\right]-\frac{l}{r}\frac{1-2p}{2}\ln\frac{1-p}{p}.\label{eq:entro-free-LDGM}
\end{equation}

\subsection{Transmission with LDPC Codes.}

We transmit code words with uniform prior, from an LDPC code with
Tanner graph $\Gamma$, over a BSC with flip probability $p$. Here
$n>m$ and $l<r$ so that the design rate $1-l/r$ is well defined.
We assume without loss of generality that the all zero codeword is
transmitted. Then the posterior probability that $\underline{x}=(x_{i})_{i=1}^{n}\in\{0,1\}^{n}$
is the transmitted word given that $\underline{y}=(y_{i})_{i=1}^{n}\in\{0,1\}^{n}$
is received, reads 
\begin{equation}
p_{\underline{X}|\underline{Y}}\left(\underline{x}|\underline{y}\right)=\frac{1}{Z_{{\rm LDPC}}}\prod_{a\in C}\mathbb{I}\left(\oplus_{i\in\partial a}x_{i}=0\right)\prod_{i\in V}\exp((-1)^{x_{i}}h_{i})\,.\label{eq:ldpc_dist}
\end{equation}
In this formula 
\begin{equation}
h_{i}=(-1)^{y_{i}}\frac{1}{2}\ln\frac{1-p}{p}
\end{equation}
are the half-log-likelihood variables and the normalizing factor $Z$
is the partition function 
\begin{equation}
Z_{{\rm LDPC}}=\sum_{\underline{x}\in\{0,1\}^{n}}\prod_{a\in C}\mathbb{I}\left(\oplus_{i\in\partial a}x_{i}=0\right)\prod_{i\in V}\exp((-1)^{x_{i}}h_{i}).\label{eq:partitionLDPC}
\end{equation}
As before the amplitude of $h_{i}$ is set to $\vert h_{i}\vert\equiv h\equiv\frac{1}{2}\ln\frac{1-p}{p}$
and the high-noise regime - $p$ close to $1/2$ - corresponds to
small $h$. The distribution of the i.i.d half-log-likelihood variables
is $c(h_{i})=(1-p)\delta(h_{i}-h)+p\delta(h_{i}+h)$. 
\begin{rem}
Equ. \eqref{eq:partitionLDPC} is the partition function of a spin
system with two types of coupling constants $\beta J_{I}\to h_{i}$
and $+\infty$. The infinite coupling constant mimics the parity check
constraints, so the high-temperature condition \ref{eq:hightemperature}
is never satisfied which makes the ensuing analysis more challenging. 
\end{rem}
The Shannon conditional entropy $H_{{\rm LDPC}}(\underline{X}\vert\underline{Y})$
of the input word given the output word $y$ is again directly related
to the free energy 
\begin{equation}
f_{{\rm LDPC}}=\frac{1}{n}\ln Z_{{\rm LDPC}}
\end{equation}
through the formula 
\begin{equation}
\frac{1}{n}H_{{\rm LDPC}}\left(\underline{X}|\underline{Y}\right)=\mathbb{E}_{\underline{h}}[f_{{\rm LDPC}}]-\frac{1-2p}{2}\ln\frac{1-p}{p}.\label{eq:entro-free}
\end{equation}

\subsection{The Bethe Approximation}

The Bethe-Peierls (mean field) theory allows one to compute candidate
approximations, called Bethe free energies, for $f=\frac{1}{n}\ln Z$.
In the case of LDPC and LDGM it allows us to compute candidate approximations
for the free energies $f_{{\rm LDPC}}$ and $f_{{\rm LDGM}}$. As
explained in the introduction, controlling in a rather systematic
way the quality of these approximation is the object of this paper.

Let us first recall the Bethe formulas for general factor graph models.
This involves a set of messages $\zeta_{i\to a}$ and $\widehat{\zeta}_{a\to i}$
attached to the edges of $(ia)\in E$. The collection of all messages
is denoted $(\underline{\zeta},\widehat{\underline{\zeta}})$; they
satisfy the belief propagation fixed point equations 
\begin{align}
\begin{cases}
\zeta_{i\rightarrow a}=\sum_{b\in\partial i\backslash a}{\widehat{\zeta}}_{b\rightarrow i}\\
\\
\tanh{\widehat{\zeta}}_{a\rightarrow i}=\frac{\sum_{s_{\partial a}}s_{i}\psi_{a}(s_{\partial a})\prod_{j\in\partial a\backslash i}(1+s_{j}\tanh\zeta_{j\to a})}{\sum_{s_{\partial a}}\psi_{a}(s_{\partial a})\prod_{j\in\partial a\backslash i}(1+s_{j}\tanh\zeta_{j\to a})}.
\end{cases}
\end{align}
The Bethe free energy associated with a particular solution of these
equations is 
\begin{equation}
f^{{\rm Bethe}}(\underline{\zeta},\underline{\widehat{\zeta}})=\frac{1}{n}\biggl(\sum_{a\in C}F_{a}+\sum_{i\in V}F_{i}-\sum_{\left(i,a\right)\in E}F_{ia}\biggr),\label{eq:fbethegeneral}
\end{equation}
where 
\begin{align}
\begin{cases}
F_{a}=\ln\bigl\{\sum_{s_{\partial a}}\psi_{a}(s_{\partial a})\prod_{j\in\partial a}(\frac{1+s_{j}\tanh\zeta_{j\to a}}{2})\bigr\},\\
\\
F_{i}=\ln\{\prod_{a\in\partial i}(1+\tanh\widehat{\zeta}_{a\rightarrow i})\\
\hskip3cm+\prod_{a\in\partial i}(1-\tanh\widehat{\zeta}_{a\rightarrow i})\},\\
\\
F_{ia}=\ln\{1+\tanh\zeta_{i\rightarrow a}\tanh\widehat{\zeta}_{a\rightarrow i}\}.
\end{cases}
\end{align}
It is easy to check that the stationary points of $f^{{\rm Bethe}}(\underline{\zeta},\underline{\widehat{\zeta}})$
considered as a function of the messages over $\mathbb{R}^{E}\times\mathbb{R}^{E}$
satisfy the Belief propagation equations.

It is immediate to specialize these formulas to LDGM codes. This yields
\begin{align}
\begin{cases}
\zeta_{i\rightarrow a}=\sum_{b\in\partial i\backslash a}{\widehat{\zeta}}_{b\rightarrow i}\\
\\
{\widehat{\zeta}}_{a\rightarrow i}=\tanh^{-1}\bigl(\tanh h_{a}\prod_{j\in\partial a\setminus i}\tanh\zeta_{j\rightarrow a}\bigr).
\end{cases}
\end{align}
and 
\begin{align}
\begin{cases}
F_{a}=\ln\{1+\tanh h_{a}\prod_{i\in\partial a}\tanh\zeta_{i\rightarrow a}\}+\ln\cosh h,\\
\\
F_{i}=\ln\{\prod_{a\in\partial i}(1+\tanh\zeta_{a\rightarrow i})\\
\hskip3cm+\prod_{a\in\partial i}(1-\tanh\zeta_{a\rightarrow i})\},\\
\\
F_{ia}=\ln\{1+\tanh\zeta_{i\rightarrow a}\tanh\widehat{\zeta}_{a\rightarrow i}\}.
\end{cases}
\end{align}
The Bethe free energy given by a sum of these three type of quantities
and is denoted by $f_{{\rm LDGM}}^{{\rm Bethe}}(\underline{\zeta},\widehat{\underline{\zeta}})$.

Since LDPC codes will require a separate treatment, in order to avoid
confusions, the messages are denoted $(\underline{\eta},\underline{\widehat{\eta}})$.
They satisfy the belief propagation fixed point equations 
\begin{align}
\begin{cases}
\eta_{i\rightarrow a}=h_{i}+\sum_{b\in\partial i\backslash a}\widehat{\eta}_{b\rightarrow i}\\
\\
\widehat{\eta}_{a\rightarrow i}=\tanh^{-1}\bigl(\prod_{j\in\partial a\setminus i}\tanh\eta_{j\rightarrow a}\bigr).\label{eq:bp_{e}quations}
\end{cases}
\end{align}
The Bethe free energy associated with a solution is 
\begin{equation}
f_{\mathrm{LDPC}}^{{\rm Bethe}}(\underline{\eta},\underline{\widehat{\eta}})=\frac{1}{n}\biggl(\sum_{a\in C}P_{a}+\sum_{i\in V}P_{i}-\sum_{\left(i,a\right)\in E}P_{ia}\biggr),\label{eq:fbethe}
\end{equation}
where 
\begin{align}
\begin{cases}
P_{a}=\ln\{1+\prod_{i\in\partial a}\tanh\eta_{i\rightarrow a}\}-\ln2,\\
\\
P_{i}=\ln\{e^{h_{i}}\prod_{a\in\partial i}(1+\tanh\eta_{a\rightarrow i})\\
\hskip3cm+e^{-h_{i}}\prod_{a\in\partial i}(1-\tanh\eta_{a\rightarrow i})\},\\
\\
P_{ia}=\ln\{1+\tanh\eta_{i\rightarrow a}\tanh\widehat{\eta}_{a\rightarrow i}\}.
\end{cases}
\end{align}
Considering the expression on the right hand side of \eqref{eq:fbethe}
as a function of $\left(\underline{\eta},\underline{\widehat{\eta}}\right)\in\mathbb{R}^{E}\times\mathbb{R}^{E}$,
allows one to check that its stationary points are solutions of the
belief propagation equations.

\section{Main Results\label{sec:main results}}

Our main interest is on standard LDPC codes Sec. \ref{sec:III.C} below. However it is useful to first consider 
``high temperature'' models for which the convergence criterion of the polymer expansion 
is much easier to assess. Two quick applications are presented in Sec. \ref{sec:III.A} and \ref{sec:III.B}.
These simple cases will allow us to see how the formalism of loop sum combined with a further polymer expansion works out 
(see Sections  \ref{sec:high temperature models}, \ref{sec:analysis for ldgm codes}).
For LDPC codes, as we will see, only part of the polymer expansion converges which makes the analysis more challenging.

One word about notation: in order to avoid adding subscripts it is understood
that we write $\lim_{n\to +\infty} A$ the quantity $A$ means a sequence $A_n$.

\subsection{Factor Graphs with Large Girth}\label{sec:III.A}

We begin with the high-temperature regime of general factor graph
models. It has been proven \cite{mooij2007sufficient} that when the
high-temperature condition \eqref{eq:hightemperature} is satisfied
the belief propagation equations have a unique fixed point solution.
In the next theorem the Bethe free energy is computed at this fixed
point. 
\begin{thm}
\label{th:hightemptheorem} Let $\Gamma_{n}$ be a sequence of Tanner
graphs, with uniformly bounded degrees and, with large girth in the
sense that ${\rm girth}(\Gamma_{n})\geq C\ln\vert\Gamma_{n}\vert$
where $C>0$ is a constant independent of $n$. Consider free energy
sequences of models constructed on $\Gamma_{n}$. For $0<\beta<\beta_{0}$
small enough we have 
\begin{equation}
\lim_{n\to+\infty}\vert f-f^{{\rm Bethe}}(\underline{\zeta},\underline{\widehat{\zeta}})\vert=0\,.\label{eq:difference}
\end{equation}
\end{thm}
\begin{rem}
Even if the individual limits of $f$ and $f^{{\rm Bethe}}$ are not
well defined their difference tends to zero. As will be seen in the
proof, the order of magnitude of this difference is $O((c\beta)^{2{\rm girth}/(2+r_{\max})})$
with $c>0$ a constant depending only on the degrees of the nodes
and the couplings $J_{I}$. 
\end{rem}

\begin{rem}
 This result could also be obtained in a different way.
 The high temperature condition implies a decay of correlations condition on the computation tree, which combined
 with the large girth condition, implies exactness of the Bethe free energy. 
\end{rem}

\subsection{LDGM Ensembles}\label{sec:III.B}

For $h$ small enough, an instance of an LDGM code is a high-temperature
graphical model with a special form of the factor weights. If the
LDGM code contains no degree one check nodes then the unique fixed
point of the belief-propagation equations is trivial i.e. $(\underline{\zeta},\underline{\widehat{\zeta}})=(\underline{0},\underline{0})$.
However if there is a non-vanishing fraction of degree one check nodes
the fixed point $(\underline{\zeta},\underline{\widehat{\zeta}})\neq (0,0)$
is non-trivial. 
\begin{thm}
\label{th:randomprop} Suppose that we draw $\Gamma$ uniformly at
random from the ensemble LDGM$(\Lambda,P,n)$. For $h<h_{0}$ small
enough we have 
\begin{equation}
\lim_{n\to+\infty}\mathbb{E}_{\Gamma}\biggl[\bigl\vert f_{{\rm LDGM}}-f_{{\rm LDGM}}^{{\rm Bethe}}(\underline{\zeta},\widehat{\underline{\zeta}})\bigr\vert\biggr]=0\,.
\end{equation}
\end{thm}

\begin{rem}
Our analysis yields $h\leq\left(4l_{\max}^{2}r_{\max}\right)^{-1}$ for the bound on the noise for LDGM codes.
For example for regular $(3,6)$ LDGM code the probability of error $p$ should
be bigger than $0.4889$, 
see \cite[p. 78]{Vuffray2014} for the details.
This can be compared to
Dobrushin's uniqueness condition applied to LDGM codes $h\leq\left(2l_{\max}r_{\max}\right)^{-1}$. 
The later condition also implies mixing {\it on the computation tree}. This type of result
has been shown in detail in \cite{Tatikonda-Jordan} for high temperature models with pairwise interactions, 
and can be extended general interactions (as already argued in \cite{Tatikonda-Jordan}).
One can check that both bounds are not very good when compared with the true phase transition threshold, 
so the present methods are not practical for determining an estimate of the true phase transition point.
This is commonly the case for such methods.
It might be thought that it is possible to improve the $l_{\max}^2$ down to $l_{\max}$ (as in Dobrushin's bound)
but this is presumably not the case at least from the present techniques. 
Indeed the different dependence here as a function of the variable node degree also occurs 
in the standard Ising model (see for example \cite{Simon} Chap V)
In general the Dobrushin uniqueness condition is weaker than analyticity conditions obtained by cluster expansions.
\end{rem}

\subsection{LDPC Ensembles}\label{sec:III.C}

Let us now describe our main result which is the analogous theorem
for LDPC codes. We assume that for $h<h_{*}$ small enough independent
of $n$ and $\epsilon>0$ independent of $n$ and $h$, there exist
a high-noise solution $(\underline{\eta},\widehat{\underline{\eta}})$
of the belief propagation equations which satisfies (see Appendix
\ref{app:activity bounds}) 
\begin{align}
\vert\tanh\eta_{i\rightarrow a}\vert\leq\left(1+\epsilon\right)\tanh h.\label{eq:high-noise}
\end{align}
The analysis does not require the uniqueness of this solution but only
its existence. We call such solutions ``high-noise solutions''.
For simplicity of notations we use throughout the analysis the quantity
$\theta$ define as
\begin{equation}
\theta=\left(1+\epsilon\right)\tanh h.\label{eq:high-noise parameter}
\end{equation}

\begin{thm}
\label{th:theorem1} Suppose $l$ is odd and $3\leq l\leq r$. There
exists $\theta_{0}>0$ (small) independent of $n$, such that for
$\theta\leq\theta_{0}$ and any high-noise solution 
\begin{equation}
\mathbb{E}_{\Gamma}\biggl[\bigl\vert\frac{1}{n}\ln Z-f_{{\rm LDPC}}^{\mathrm{Bethe}}\left(\underline{\eta},\underline{\widehat{\eta}}\right)\bigr\vert\biggl]=O\bigl(n^{-(l(1-\kappa)-1)}\bigr)\,.\label{eq:first-result}
\end{equation}
The $O(\cdot)$ is uniform in the channel output realizations $\underline{h}$. \end{thm}
\begin{rem}
We recall that $\kappa\in]1-\frac{2(r-1)}{lr},1-\frac{1}{l}[$ which
implies that the expansion constant $\kappa$ is such that, for $r>2$,
$0<l(1-\kappa)-1<(r-2)/r$. 
\end{rem}

\begin{rem}
 To the best of our knowledge this type of result for LDPC codes has never been obtained before. Related but different results 
 in \cite{kudekar2011decay}, already alluded to in the introduction concern the low noise regime. Moreover as remarked there, the 
 Dobrushin uniqueness condition fails for LDPC codes and the situation is qualitatively different than in the 
 high temperature or LDGM cases.
\end{rem}

\section{Loop Sum and Polymer Expansion\label{sec:loop sum and polymer expansion}}

The formalism developed in this section is valid for general graphical
models, and in particular for fixed instances of LDPC and LDGM codes.
We give only the necessary information needed for the subsequent analysis
in Sections \ref{sec:high temperature models}, \ref{sec:analysis for ldgm codes},
\ref{sec: analysis for ldpc codes}. More details can be found in
appendices \ref{app:loop sum identity}, \ref{app:polymer}.

\subsection{Polymer Representation}

\begin{figure}[ptb]
\centering{}\includegraphics[scale=0.19]{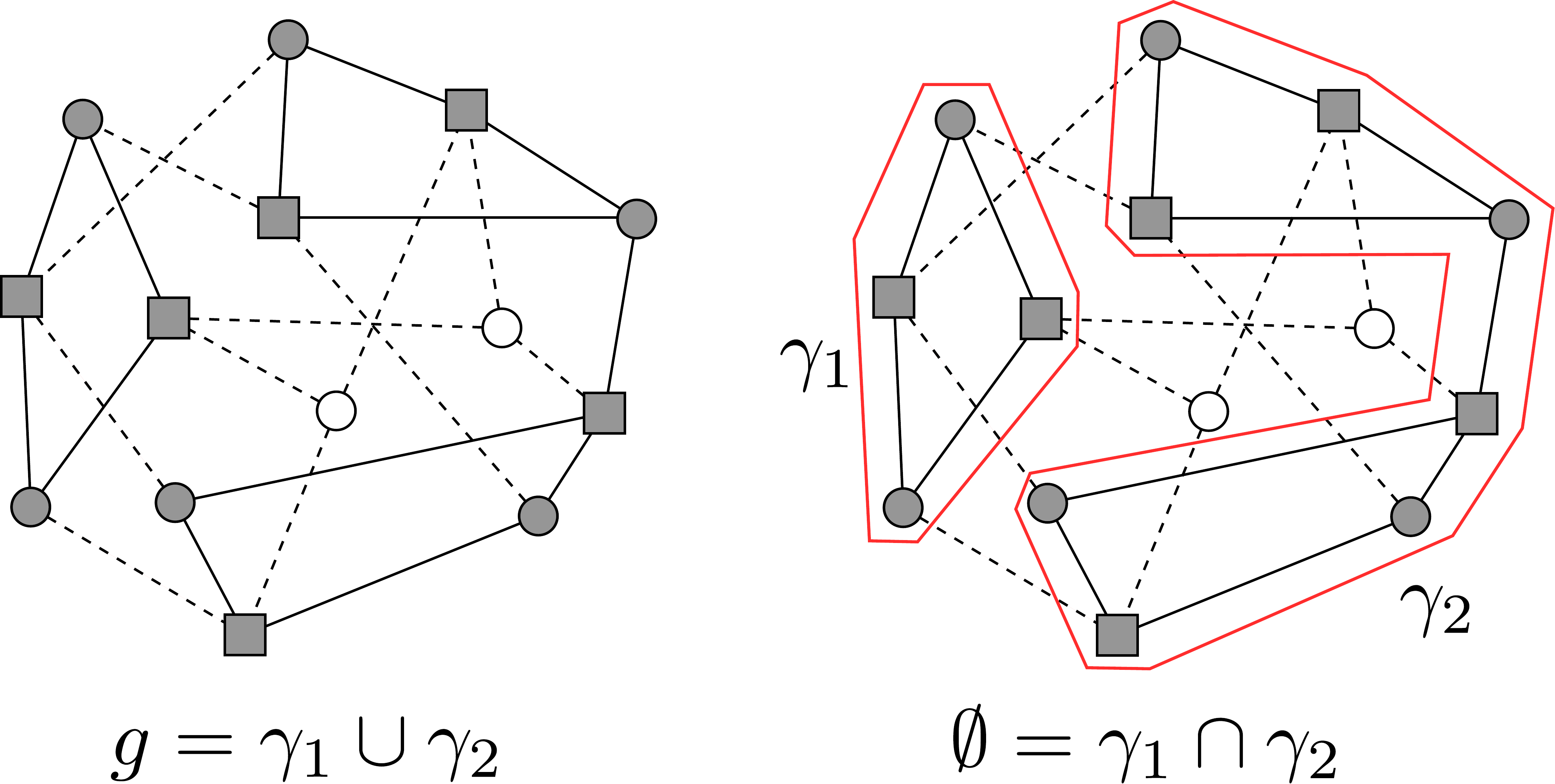} \protect\caption{\label{fig:polymers} Example of $\Gamma\in\mathcal{B}\left(3,4,8\right)$.
The generalized loop $g$ has two disjoint connected parts $\gamma_{1}$
and $\gamma_{2}$.}
\end{figure}

Take a subset of edges of $\Gamma$ together with the end-vertices
of these edges. This forms a subgraph $g$ of $\Gamma$. We call $d_{i}(g)$
(resp. $d_{a}(g)$) the induced degree of node $i$ (resp. $a$) in
a subgraph $g$. If $d_{i}\left(g\right)\geq2$ and $d_{a}\left(g\right)\geq2$
for all $i,a\in g$, we say that $g$ is a loop. In other words a
loop has no dangling edge. Note that a loop is not necessarily a cycle,
and is not necessarily connected. Figure \ref{fig:polymers} shows
an example.

For a finite size system, Chertkov and Cherniak \cite{chertkov2006loop}
derived the following loop-sum identity 
\begin{equation}
Z=\exp(nf^{\mathrm{Bethe}})\times\bigl(1+\sum_{g\subset\Gamma}K\left(g\right)\bigr).\label{eq:cc-identity}
\end{equation}
where each quantity on the right hand side is computed for a solution
of the belief propagation equations. The sum on the right hand side
carries over all loops included in $\Gamma$. As long as the graph
is finite, this is a finite sum which is well defined. The quantities
$K(g)$ can be expressed entirely in terms of belief propagation messages
$(\zeta_{i\to a},\widehat{\zeta}_{a\to i})$ or $(\eta_{i\to a},\widehat{\eta}_{a\to i})$
such that $i$ or $a$ belong to $g$. The explicit formulas for general
models as well as for LDPC and LDGM codes are given in Appendix \ref{app:activity bounds}.
For the convenience of the reader we give a short derivation of identity
\eqref{eq:cc-identity} in Appendix \ref{app:loop sum identity}.

Each generalized loop can be decomposed in a unique way as an union
of its connected components
\begin{equation}
g=\cup_{k}\gamma_{k},\label{eq:polymer decomposition}
\end{equation}
where $\gamma_{k}$ are the connected components of $g$. It is easy
to see that the $\gamma_{k}$ entering in \eqref{eq:polymer decomposition}
are non-empty connected loops and are mutually disjoint (see Figure
\ref{fig:polymers}). The connected loops $\gamma_{k}$ are called
polymers. Remarkably each $K(g)$ can be factorized (see Appendix
\ref{app:loop sum identity}, eqn. \eqref{eq:total_activity}) in
a product of contributions associated with the connected parts of
$g$. We have\footnote{We note that this factorization is not necessarily unique and in practice
one should choose the most natural one.} 
\begin{equation}
K(g)=\prod_{k}K(\gamma_{k}).\label{eq:factorization}
\end{equation}
The factorization implies 
\begin{equation}
1+\sum_{g\subset\Gamma}K\left(g\right)\equiv Z^{{\rm polymer}},
\end{equation}
with 
\begin{align}
Z^{{\rm polymer}}=\sum_{M\geq0} & \frac{1}{M!}\sum_{\gamma_{1},...,\gamma_{M}\subset\Gamma}\prod_{k=1}^{M}K\left(\gamma_{k}\right)\nonumber \\
 & \times\prod_{k<k^{\prime}}\mathbb{I}\left(\gamma_{k}\cap\gamma_{k^{\prime}}=\emptyset\right).\label{eq: z polymer}
\end{align}
In the second sum on the right hand side, each $\gamma_{k}$ runs
over all polymers contained in $\Gamma$. The factor $\frac{1}{M!}$
accounts for the fact that a polymer configuration has to be counted
only once. The indicator function ensures that the polymers do not
intersect. By convention the term $M=0$ is equal to $1$ and for
$M=1$ the indicator function equals $1$. Note that because of the
non-intersection constraint of the polymers, the number of terms in
the sums on the right hand side is finite.

From a physical point of view \eqref{eq: z polymer} interprets the
loop sum in Equ. \eqref{eq:cc-identity} as the \textit{partition
function of a gas of polymers} that can acquire any shape allowed
by $\Gamma$, have \textit{activity}\footnote{``Activity'' is the name used by chemists for the prior probability
weight $K(\gamma)$ of an isolated polymer. Note that here $K(\gamma)$
can be negative and this analogy is at best formal. We use the name
``activity'' rather than ``prior weight'' for $K(\gamma)$ precisely
because they can be negative in the present context.} $K(\gamma)$, and interact via a two body hard-core repulsion which
precludes their overlap. This analogy allows us to use methods from
statistical mechanics to analyze the corrections to the Bethe free
energy.

\subsection{Polymer Expansion}

All the corrections to the Bethe free energy are contained in the
free energy of the polymer gas, namely 
\begin{equation}
\frac{1}{n}\ln Z=f^{\mathrm{Bethe}}+\frac{1}{n}\ln Z^{{\rm polymer}}.\label{eq:polfree}
\end{equation}

We start with a heuristic discussion in order to motivate the ensuing
formalism. If a suitable fixed point of the belief propagation equations
is chosen such that the Bethe free energy is a good approximation,
then we expect that the polymer free energy is small (or negligible
in an appropriate limit). One way that this may happen is if the activities
of the polymers become small as the size of the polymers increase.
Let us explain this point in more detail. We expect the activities
to be exponentially small in $\vert\gamma\vert$ (as will become clear
later for LDPC and LDGM this is true for small $h$). This smallness
of the activities is counterbalanced by an entropic contribution that
accounts for the large number of polymers of given size. This number
is exponentially large in $\vert\gamma\vert$. For $h$ small enough
the smallness of the activities wins over the entropic terms and one
can expand the $\log$ in a power series in $K(\gamma)$. Since the
polymers have no dangling edges, on a locally tree like graph they
have a typical size $\vert\gamma\vert\approx c\ln n$ for some small
constant $c$. This means $K(\gamma)\approx O(e^{-c\ln n})$ and since
the series expansion starts linearly with $K(\gamma)$, the polymer
free energy is itself $O(e^{-c\ln n})$. Note that the polymer free
energy could still be negligible even if the activities are not small
because in general they have signs and cancellations could occur.
However such cancellations would be difficult to control. The regimes
investigated in this paper are those where the activities are small
enough so that their weight counterbalances the entropy of the polymers
and we do not need to track sign cancellations.

As just explained, with small activities it is natural to expand the
logarithm on the right hand side of \eqref{eq:polfree} and to control
the convergence of this expansion. Such expansions are called { polymer
expansions}\footnote{They are also called Mayer expansions. another more generic name is
cluster expansion.}. We now describe the combinatorial structure of the polymer expansion
and explain what convergence criteria are available.

\begin{figure}[tbh]
\centering{}\includegraphics[scale=0.5]{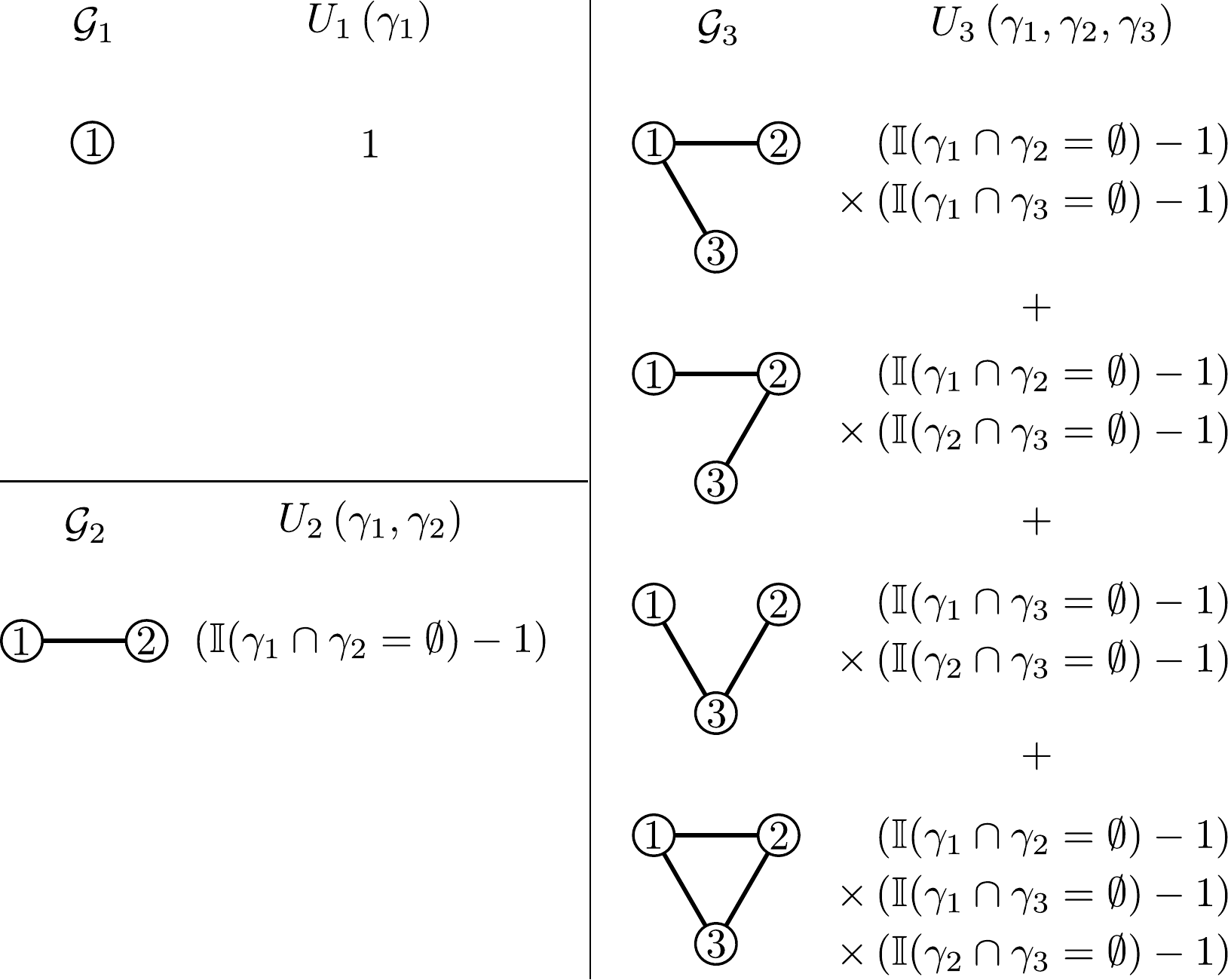} \protect\caption{\label{fig:mayer graph} All the Mayer graphs for $M=1,2,3$ and their
associated Ursell functions.}
\end{figure}

We introduce the set $\mathcal{G}_{M}$ of all connected graphs $G$
with $M$ labeled vertices $1,\cdots,M$ (see Figure \ref{fig:mayer graph}).
These are called Mayer graphs. We associate to an ensemble of Mayer
graphs $\mathcal{G}_{M}$ an Ursell function whose arguments are polymers
\begin{equation}
U_{M}(\gamma_{1},\ldots,\gamma_{M})=\sum_{G\in\mathcal{G}_{M}}\prod_{(k,k^{\prime})\in G}\left(\mathbb{I}\left(\gamma_{k}\cap\gamma_{k^{\prime}}=\emptyset\right)-1\right),
\end{equation}
were the notation $\prod_{(k,k^{\prime})\in G}$ denotes a product
over the edges $\left(k,k'\right)$ of $G$. By convention $U_{M=1}\left(\gamma_{1}\right)\equiv1$.
Notice that an Ursell function is equal to zero if the polymers are
two-by.two disjoint. Expanding the logarithm in the free energy of
polymers in powers of the activities yields the expansion 
\begin{eqnarray}
\frac{1}{n}\ln Z^{{\rm polymer}} & = & \frac{1}{n}\sum_{M=1}^{+\infty}\frac{1}{M!}\sum_{\gamma_{1},...,\gamma_{M}\subset\Gamma}\prod_{k=1}^{M}K\left(\gamma_{k}\right)\nonumber \\
 &  & \times U_{M}\left(\gamma_{1},\cdots,\gamma_{M}\right).\label{eq:polymerexpansion}
\end{eqnarray}
A short check of this identity is given in Appendix \ref{app:polymer}.

It is important to note that now the first two sums on the right hand
side are infinite because the Ursell functions force polymers to overlap.
It is therefore important to control the convergence of this formal
power series. A standard criterion for uniform (with respect to $n$)
convergence is that 
\begin{equation}
Q\equiv\sum_{t=0}^{\infty}\frac{1}{t!}\sup_{x\in V\cup C}\sum_{\gamma\ni x}\left\vert \gamma\right\vert ^{t}\left\vert K\left(\gamma\right)\right\vert <1\,,\label{eq:convcrit}
\end{equation}
where the last sum in runs over polymers $\gamma$ containing $x$.
This implies in particular that the polymer free energy is analytic
as a function of $\{K(\gamma),\gamma\subset\Gamma\}$.

A mathematically precise and simple way to express the analyticity
of the series is to replace $K(\gamma)$ by $zK(\gamma)$, $z\in\mathbb{C}$,
$\vert z\vert<z_{0}$, where $z_{0}>1$ is fixed. Then the polymer
free energy becomes a function of the complex variable $z$, 
\begin{equation}
\frac{1}{n}\ln Z^{{\rm polymer}}(z)\label{eq:zquantity}
\end{equation}
and \eqref{eq:polymerexpansion} becomes a series expansion in $z^{M}$,
$M\geq1$. If the convergence criterion \eqref{eq:convcrit} holds
with $K(\gamma)$ replaced by $z_{0}K(\gamma)$ we can conclude that
the series is holomorphic for $\vert z\vert<z_{0}$. Moreover the
limit $n\to+\infty$, as long as it exists, is also holomorphic for
$\vert z\vert<z_{0}$. In practice, existence of the limit requires
some regularity structure on the sequence of graphical models (which
is not the case in the present formulation), and it can be checked
term by term on the series expansion. We take $z_{0}>1$ in order
to then apply the results to the case of interest $z=1$.

As will be seen in Sections \ref{sec:high temperature models}, \ref{sec:analysis for ldgm codes}
it is fairly easy to check that \eqref{eq:convcrit} is satisfied
for high-temperature general models and also for typical instances
of LDGM codes in the large-noise regime. This case also serves as
a pedagogical one to better understand the difficulties that arise
in the case of LDPC codes. In fact for LDPC codes we are not able
to satisfy this criterion as such. However the criterion holds if
$\Gamma$ is an expander and the sum in $Z^{{\rm polymer}}$ is restricted
to { small} polymers of size $\vert\gamma\vert<\lambda n$, $0<\lambda<\lambda_{0}$
(recall $\lambda_{0}$ is defined in Section \ref{subsec:factor graphs}).
The contribution of ``large'' polymers $\vert\gamma\vert>\lambda n$
is treated differently.

\section{High-Temperature Models\label{sec:high temperature models}}

We recall that when the high-temperature condition \eqref{eq:hightemperature}
is satisfied the fixed point solution of the belief propagation equations
is unique \cite{mooij2007sufficient}. Moreover we show in Appendix
\ref{app:activity bounds} that it satisfies 
\begin{equation}
\vert\tanh\zeta_{i\to a}\vert\leq2(l_{{\rm max}}-1)\mu,\,\,\,\,\vert\tanh\widehat{\zeta}_{a\to i}\vert\leq2\mu,\label{eq:fphightemp}
\end{equation}
where $\mu$ is define by Equation \eqref{eq:hightemperature} and
is proportional to the temperature $\beta$.
\begin{lem}
\label{th:lemmahightemp} Consider the $z$ dependent free energy
defined in \eqref{eq:zquantity} computed at the fixed point \eqref{eq:fphightemp}.
One can find a $\beta_{0}>0$ small enough such that for $0<\beta<\beta_{0}$,
such that: 
\begin{enumerate}
\item $n^{-1}\ln Z^{{\rm polymer}}(z)$ has an absolutely uniformly (in
$n$) convergent power series expansion in $z^{M}$, $M\geq1$ for
$\vert z\vert<z_{0}$. 
\item If one considers a sequence of factor graph models such that the thermodynamic
limit $\lim_{n\to+\infty}n^{-1}\ln Z^{{\rm polymer}}(z)$ exists,
this limit is an analytic function of $z$ for $\vert z\vert<z_{0}$. 
\item 
\begin{equation}
\frac{1}{n}\vert\ln Z^{{\rm polymer}}(z)\vert\leq\frac{4}{n}z_{0}\sum_{x\in V\cup C}\sum_{\gamma\ni x}(6e\mu)^{\frac{2\vert\gamma\vert}{2+r_{\max}}}e^{\vert\gamma\vert}\,.\label{eq:ineg sur f}
\end{equation}

\end{enumerate}
\end{lem}
\begin{rem}
Note that the second statement is an immediate consequence of the
first one. Later we make use of the third statement for $z=1$.\end{rem}
\begin{IEEEproof}
For the activities of the polymers computed at the fixed point we
have the bounds (Appendix \ref{app:activity bounds}) 
\begin{equation}
\vert K(\gamma)\vert\leq(6e\mu)^{\frac{2\vert\gamma\vert}{2+r_{{\rm max}}}}.\label{eq:acti}
\end{equation}
Next we use the remarkable inequality \cite{brydges1984short} 
\begin{align}
\left|U_{M}\left(\gamma_{1},\cdots,\gamma_{M}\right)\right|\leq & \sum_{T\in\mathcal{T}_{M}}\prod_{\left(k,k'\right)\in T}\vert\mathbb{I}\left(\gamma_{k}\cap\gamma_{k'}=\emptyset\right)-1\vert\nonumber \\
= & \sum_{T\in\mathcal{T}_{M}}\prod_{\left(k,k'\right)\in T}\mathbb{I}\left(\gamma_{k}\cap\gamma_{k'}\neq\emptyset\right),\label{eq:bryd}
\end{align}
where $\mathcal{T}_{M}$ is the set of trees on $M$ vertices labeled
$1,\cdots,M$. Using \eqref{eq:acti} and \eqref{eq:bryd} we find
that the term of order $M$ in \eqref{eq:polymerexpansion} is smaller
than 
\begin{equation}
\frac{1}{M!}\sum_{T\in\mathcal{T}_{M}}\sum_{\gamma_{1},...,\gamma_{M}}\prod_{\left(k,k'\right)\in T}\mathbb{I}\left(\gamma_{k}\cap\gamma_{k'}\neq\emptyset\right)\prod_{k=1}^{M}z_{0}(6e\mu)^{\frac{2\vert\gamma_{k}\vert}{2+r_{\max}}}.\label{eq:bound tree graph}
\end{equation}
We will now estimate the sum over $\gamma_{1},...,\gamma_{M}$ for
each tree $T$. Let $t_{1},...,t_{M}$ be the degrees of the nodes
$1,...,M$. One can decide that $\gamma_{1}$ labels the root of $T$
and that the leafs are among $2,...,M$. We first perform recursively
the sum over $\gamma_{2},...,\gamma_{M}$ by starting from the leaf
nodes in this set. One finds the estimate 
\begin{align}
\sum_{\gamma_{2},...,\gamma_{M}} & \prod_{k=2}^{M}z_{0}(6e\mu)^{\frac{2\vert\gamma_{k}\vert}{2+r_{\max}}}\prod_{\left(k,k'\right)\in T\setminus\{1\}}\mathbb{I}\left(\gamma_{k}\cap\gamma_{k'}\neq\emptyset\right)\nonumber \\
 & \leq\vert\gamma_{1}\vert^{t_{1}}\prod_{k=2}^{M}\biggl\{\sup_{x\in V\cup C}\sum_{\gamma\ni x}\vert\gamma\vert^{t_{k}-1}z_{0}(6e\mu)^{\frac{2\vert\gamma\vert}{2+r_{\max}}}\biggr\}.\label{eq:bound leaf nodes}
\end{align}
This implies 
\begin{align}
\sum_{\gamma_{1},...,\gamma_{M}} & \prod_{k=1}^{M}z_{0}(6e\mu)^{\frac{2\vert\gamma_{k}\vert}{2+r_{\max}}}\prod_{\left(k,k'\right)\in T}\mathbb{I}\left(\gamma_{k}\cap\gamma_{k'}\neq\emptyset\right)\nonumber \\
 & \leq\sum_{y\in V\cup C}\sum_{\gamma_{1}\ni y}z_{0}(6e\mu)^{\frac{2\vert\gamma_{1}\vert}{2+r_{\max}}}\vert\gamma_{1}\vert^{t_{1}}\nonumber \\
 & \times\prod_{k=2}^{M}\biggl\{\sup_{x\in V\cup C}\sum_{\gamma\ni x}\vert\gamma\vert^{t_{k}-1}z_{0}(6e\mu)^{\frac{2\vert\gamma\vert}{2+r_{\max}}}\biggr\}.\label{eq:second bound leaf nodes}
\end{align}
Now it is easy to estimate the sum over $T$ in \eqref{eq:bound tree graph}.
According to the Cayley formula the number of trees with $M$ vertices
of degrees $t_{1},...,t_{M}$ is equal to 
\begin{equation}
\frac{(M-2)!}{(t_{1}-1)!...(t_{M}-1)!},
\end{equation}
so we find that \eqref{eq:bound tree graph} is upper bounded by 
\begin{align}
\sum_{y\in V\cup C} & \sum_{\gamma_{1}\ni y}z_{0}(6e\mu)^{\frac{2\vert\gamma_{1}\vert}{2+r_{\max}}}e^{\vert\gamma_{1}\vert}\frac{1}{M}\nonumber \\
 & \times\biggl\{\sum_{t=0}^{+\infty}\frac{1}{t!}\sup_{x\in V\cup C}\sum_{\gamma\ni x}\vert\gamma\vert^{t}z_{0}(6e\mu)^{\frac{2\vert\gamma\vert}{2+r_{\max}}}\biggr\}^{M-1}.\label{eq: final bound leaf nodes}
\end{align}
We will check that in this expression the quantity in brackets 
\begin{equation}
Q^{*}\equiv\sum_{t=0}^{+\infty}\frac{1}{t!}\sup_{x\in V\cup C}\sum_{\gamma\ni x}\vert\gamma\vert^{t}z_{0}(6e\mu)^{\frac{2\vert\gamma\vert}{2+r_{\max}}}.
\end{equation}
can be made smaller than $1/2$ for $\beta_{0}$ small enough. This
implies the first statement of the lemma.

The number of polymers with size $\left|\gamma\right|=t$ containing
a node $x\in V\cup C$ is upper-bounded by $e^{At}$ where $A>0$
is a numerical constant depending only on the maximal degrees of $\Gamma$
(See Appendix \ref{app:On-the-number}). As a polymer contains at
least two nodes, one finds 
\begin{align}
Q^{*} & \leq\sum_{t=2}^{+\infty}z_{0}(6e\mu)^{\frac{2t}{2+r_{\max}}}e^{t(A+1)}\nonumber \\
 & \leq z_{0}\frac{(6e\mu)^{\frac{4}{2+r_{\max}}}e^{2(A+1)}}{1-(6e\mu)^{\frac{2}{2+r_{\max}}}e^{(A+1)}}\leq\frac{1}{2}
\end{align}
Summing \eqref{eq: final bound leaf nodes} over $M\geq1$ and using
$Q^{*}\leq1/2$ yields the third statement \eqref{eq:ineg sur f}. 
\end{IEEEproof}
An immediate application of this lemma is the proof of Theorem \ref{th:hightemptheorem}. 
\begin{IEEEproof}[\textit{Proof of Theorem \ref{th:hightemptheorem}}]
\textit{ Let $n$ the number of variable nodes of the graph $\Gamma_{n}$.
Proving Equ. \eqref{eq:difference} is equivalent to 
\begin{equation}
\lim_{n\to+\infty}\frac{1}{n}\ln Z^{{\rm polymer}}=0.
\end{equation}
To show this we will use estimate \eqref{eq:ineg sur f} for $z=1$.
The graphs $\Gamma_{n}$ have large girth, and since a polymer $\gamma\subset\Gamma_{n}$
containing $x$ certainly have at least one closed cycle, we have
$\vert\gamma\vert\geq C\ln n$ (for $C>0$ not too large). Using this
fact and that the number of such polymers is less than $e^{A\vert\gamma\vert}$
we find 
\begin{align}
\frac{1}{n}\vert\ln Z^{{\rm polymer}}\vert\leq & \frac{4}{n}\sum_{x\in\Gamma_{n}}\sum_{\gamma\ni x}(6e\mu)^{\frac{2\vert\gamma\vert}{2+r_{\max}}}e^{\vert\gamma\vert}\nonumber \\
\leq4\left(1+\frac{l}{r}\right) & \sum_{t\geq C\ln n}^{+\infty}(6e\mu)^{\frac{2t}{2+r_{\max}}}e^{t(A+1)}.\label{eq:bound polymer high temp}
\end{align}
We recall that $\mu$ is proportional to $\beta$ (see Equation \eqref{eq:hightemperature}).
Clearly there exist $\beta_{0}>0$ such that this estimate tends to
zero as $n\to+\infty$ for $\beta<\beta_{0}$. In fact we have that
$n^{-1}\ln\left|Z^{\text{poylmer}}\right|=O(\beta^{2{\rm girth}/(2+r_{\max})})$.} 
\end{IEEEproof}

\section{Analysis for LDGM Codes\label{sec:analysis for ldgm codes}}

For $h$ small enough the fixed point solution of the belief propagation
equations of an LDGM$(\Lambda,P)$ ensemble satisfies 
\begin{equation}
\vert\tanh\zeta_{i\to a}\vert\leq4(l_{{\rm max}}-1)h,\,\,\,\,\vert\tanh\widehat{\zeta}_{a\to i}\vert\leq4h.\label{eq:bound fixed-point-ldgm}
\end{equation}
For the activities of the polymers computed at the fixed point we
have the bounds (Appendix \ref{app:activity bounds}) 
\begin{equation}
\vert K(\gamma)\vert\leq\left(12eh\right)^{2\vert\gamma\vert/(2+r_{\max})}.\label{eq:act}
\end{equation}
Therefore Lemma \ref{th:lemmahightemp} applies with $\beta J$ replaced
by $h$. This allows us to prove Theorem \ref{th:randomprop}. 
\begin{IEEEproof}[\textit{Proof of Theorem \ref{th:randomprop}}]
\textit{ For $h$ small enough Lemma \ref{th:lemmahightemp} implies
\begin{equation}
\frac{1}{n}\vert\ln Z_{{\rm LDGM}}^{{\rm polymer}}\vert\leq\frac{4}{n}\sum_{x\in V\cup C}\sum_{\gamma\ni x}(12eh)^{\frac{2\vert\gamma\vert}{2+r_{{\rm max}}}}e^{\vert\gamma\vert}
\end{equation}
Taking the expectation of this inequality, 
\begin{equation}
\mathbb{E}_{\Gamma}\biggl[\frac{1}{n}\vert\ln Z_{{\rm LDGM}}^{{\rm polymer}}\vert\biggr]\leq4\left(1+\frac{l}{r}\right)\mathbb{E}_{\Gamma}\biggl[\sum_{\gamma\ni o}(12eh)^{\frac{2\vert\gamma\vert}{2+r_{{\rm max}}}}e^{\vert\gamma\vert}\biggr],\label{eq:expect bound polymer high temp}
\end{equation}
where $o\in V\cup C$ is chosen arbitrarily. Given $\Gamma$, let
$N_{R}(o)$ be the subgraph formed by the set of nodes that are at
distance less than $R$ from $o$. For the moment $R$ is a fixed
number. For $R$ fixed and $n$ large enough, this subgraph is a tree
with probability 
\begin{equation}
1-O(\frac{C_{l_{\max},r_{\max},R}}{n}),
\end{equation}
where $C_{l_{\max},r_{\max},R}>0$ depends only on $R$ and the maximal
degrees. This means that for $n$ large enough the polymers $\gamma\ni o$
have a size $\vert\gamma\vert\geq R$. Thus for $R$ fixed and $n$
large enough 
\begin{align}
\mathbb{E}_{\Gamma}\biggl[\sum_{\gamma\ni o} & (12eh)^{\frac{2\vert\gamma\vert}{2+r_{{\rm max}}}}e^{\vert\gamma\vert}\biggr]\nonumber \\
 & \leq(1-O(\frac{C_{l_{\max},r_{\max},R}}{n}))\sum_{t\geq R}((12eh)^{\frac{2}{2+r_{{\rm max}}}}e^{A+1})^{t}\nonumber \\
 & +O(\frac{C_{l_{\max},r_{\max},R}}{n})\sum_{t\geq3}((12eh)^{\frac{2}{2+r_{{\rm max}}}}e^{A+1})^{t}.
\end{align}
Replacing this estimate in \eqref{eq:expect bound polymer high temp}
and taking the limit $n\to+\infty$, 
\begin{equation}
\lim_{n\to+\infty}\mathbb{E}_{\Gamma}\biggl[\frac{1}{n}\vert\ln Z_{{\rm LDGM}}^{{\rm polymer}}\vert\biggr]\leq\sum_{t\geq R}((12eh)^{\frac{2}{2+r_{{\rm max}}}}e^{A+1})^{t}.
\end{equation}
Finally, taking the limit $R\to+\infty$ ends the proof.} 
\end{IEEEproof}

\section{Analysis for LDPC Codes\label{sec: analysis for ldpc codes}}

Recall that $\theta=\left(1+\epsilon\right)\tanh h$ (eqn. \eqref{eq:high-noise parameter}).
From \eqref{eq:high-noise} we deduce in Appendix \ref{app:activity bounds}
a (qualitatively) optimal estimate \eqref{eq:actibound}, \eqref{eq: bound activity}
on the activity of a polymer.

\subsection{Contribution of Small Polymers\label{subsec:small polymer contribution}}

\begin{figure}[ptb]
\centering{}\includegraphics[scale=0.18]{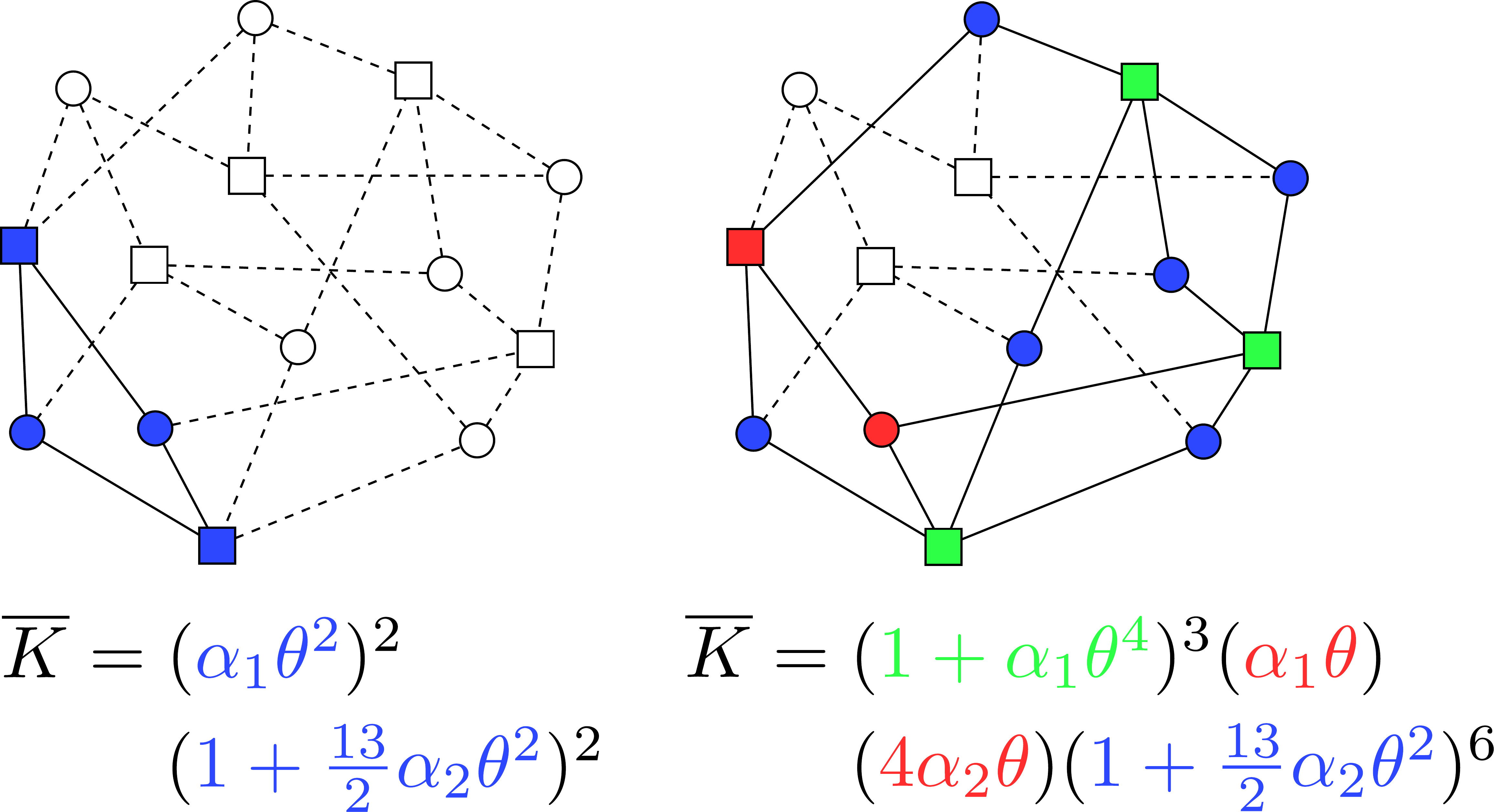}\protect\caption{\label{fig:poly_activities} Example for a $\Gamma\in\mathcal{B}\left(3,4,8\right)$
of polymers and their associated bound on their activity. The constants
$\alpha_{1}$ and $\alpha_{2}$ are close to 1 (see Appendix \ref{app:activity bounds}).
On the left a small polymer and on the right a large polymer.}
\end{figure}

The estimate in Appendix \ref{app:activity bounds} given by \eqref{eq:actibound}
and \eqref{eq: bound activity} is quite cumbersome, so let us begin
with a few remarks to understand its main qualitative features. The
activity $K(\gamma)$ is not necessarily very small for graphs containing
too many check nodes of maximal induced degree and too many variable
nodes of even induced degree. More precisely for these ``bad graphs''
the rate of decay as $\vert\gamma\vert$ grows is too slow even for
$\theta$ small, and it is not clear that it counterbalances the exponentially
large entropic terms. However the rate of decay as $\vert\partial\gamma\cap C\vert$
grows is large for $\theta$ small. Here the boundary $\partial\gamma\cap C$
is by definition the set of check nodes in $\gamma$ of non-maximal
induced degree. An example is shown on figure \ref{fig:poly_activities}.

For $\Gamma\subset\mathcal{B}(l,r,n)$ that are expanders, if $\gamma$
is ''small'' then $\vert\partial\gamma\cap C\vert$ is of the order
of $\vert\gamma\vert$ and the activity is exponentially small in
the size of the polymer. This is the meaning of the following lemma. 
\begin{lem}
\label{th:expander_activity}Assume that $\Gamma$ is a $(\lambda,\kappa)$
expander with $\kappa\in]1-\frac{2(r-1)}{lr},1-\frac{1}{l}[$. For
$\vert\gamma\vert<\lambda n$ we have for $\theta$ small enough 
\begin{equation}
\vert K(\gamma)\vert\leq\theta^{\frac{c}{2}\vert\gamma\vert},\label{eq:bound-activity}
\end{equation}
with 
\begin{equation}
c=r-\frac{2+r}{3-l(1-\kappa)}\,.\label{eq:c}
\end{equation}
\end{lem}
\begin{rem}
In the process of this derivation one has to require $3-l(1-\kappa)>0$
and $c>0$. This imposes the condition on the expansion constant $\kappa>1-\frac{2(r-1)}{lr}$.
Note that an expansion constant cannot be greater than $1-1/l$, so
it is fortunate that we have $1-\frac{1}{l}>1-\frac{2(r-1)}{lr}$
(for any $r>2$). \end{rem}
\begin{IEEEproof}
Recall that $d_{i}(\gamma)$ (resp. $d_{a}(\gamma)$) is the induced
degree of node $i$ (resp. $a$) in $\gamma$. The type of $\gamma$
is given by two vectors $\underline{n}=(n_{s}(\gamma))_{s=2}^{l}$
and $\underline{m}=(m_{t}(\gamma))_{t=2}^{r}$ defined as $n_{s}\left(\gamma\right):=\left\vert \left\{ i\in\gamma\cap V\vert d_{i}(\gamma)=s\right\} \right\vert $
and $m_{t}\left(\gamma\right):=\left\vert \left\{ a\in\gamma\cap C\vert d_{a}(\gamma)=t\right\} \right\vert $.
In words, $n_{s}(\gamma)$ and $m_{t}(\gamma)$ count the number of
variable and check nodes with induced degrees $s$ and $t$ in $\gamma$.
Note that we have the constraints 
\begin{equation}
\begin{cases}
\vert\gamma\vert=\sum_{s=2}^{l}n_{s}(\gamma)+\sum_{t=2}^{r}m_{t}(\gamma)\\
\sum_{s=2}^{l}sn_{s}(\gamma)=\sum_{t=2}^{r}tm_{t}(\gamma)
\end{cases}\label{eq:case}
\end{equation}
We apply the expander property to the set $\mathcal{V=}\left\{ i\in\gamma\cap V\right\} $.
This reads 
\begin{equation}
\vert\partial\mathcal{V}\vert\geq\kappa l\sum_{s=2}^{l}n_{s}\left(\gamma\right)\,.\label{eq:exp}
\end{equation}
On the other hand $\vert\partial\mathcal{V}\vert\leq\sum_{t=2}^{r}m_{t}(\gamma)+\sum_{s=2}^{l}(l-s)n_{s}(\gamma)$.
With \eqref{eq:case} and \eqref{eq:exp} this yields the constraint
\begin{equation}
\sum_{t=2}^{r}(r-t)m_{t}(\gamma)\geq-\vert\gamma\vert l(1-\kappa)+(l(1-\kappa)+r-1)\sum_{t=2}^{r}m_{t}(\gamma).\label{eq:first_inequality}
\end{equation}
Relaxing the second constraint in \eqref{eq:case} gives 
\begin{equation}
\sum_{t=2}^{r}tm_{t}(\gamma)\geq2\sum_{s=2}^{l}n_{s}(\gamma).
\end{equation}
Combined with the first constraint of \eqref{eq:case} this yields
\begin{equation}
(r+2)\sum_{t=2}^{r}m_{t}(\gamma)\geq2\vert\gamma\vert+\sum_{t=2}^{r}(r-t)m_{t}(\gamma).\label{eq:second_inequality}
\end{equation}
We have by use of inequalities \eqref{eq:first_inequality} and \eqref{eq:second_inequality}
\begin{equation}
\sum_{t=2}^{r-1}(r-t)m_{t}(\gamma)\geq\bigl(r-\frac{2+r}{3-l(1-\kappa)}\bigr)\vert\gamma\vert.
\end{equation}
Finally, by bounding the product over $t=2,\cdots,r-1$ in the activity
bound \eqref{eq: bound activity} of Appendix \ref{app:activity bounds},
we obtain \eqref{eq:bound-activity}. 
\end{IEEEproof}
We say that a polymer is small if $\vert\gamma\vert<\lambda n$. We
define the partition function (with activities computed at the fixed
point $(\underline{\eta},\widehat{\underline{\eta}})$ of a gas of
small polymers 
\begin{align}
Z_{{\rm small}}=\sum_{M\geq0} & \frac{1}{M!}\sum_{\gamma_{1},...,\gamma_{M}~\mathrm{s.t}~\left\vert \gamma_{k}\right\vert <\lambda n}\prod_{k=1}^{M}K\left(\gamma_{k}\right)\nonumber \\
 & \times\prod_{k<k^{\prime}}\mathbb{I}\left(\gamma_{k}\cap\gamma_{k^{\prime}}=\emptyset\right).
\end{align}
The free energy of the gas of small polymers $n^{-1}\ln Z_{{\rm small}}$
has a polymer expansion \eqref{eq:polymerexpansion} with the second
sum replaced by a sum over $\gamma_{1},\cdots,\gamma_{M}$ s.t $\vert\gamma_{i}\vert<\lambda n$. 
\begin{lem}
\label{th:convergence-lemma} Suppose $r>2$. Fix $z_{0}>1$ and replace
$K(\gamma)$ by $zK(\gamma)$, $z\in\mathbb{C}$, $\vert z\vert<z_{0}$,
in the polymer expansion of $n^{-1}\ln Z_{{\rm small}}$ which now
becomes a power series in the parameter $z^{M}$, $M\geq1$. Assume
that $\Gamma$ is a $(\lambda,\kappa)$ expander with $\kappa\in]1-\frac{2(r-1)}{lr},1-\frac{1}{l}[$.
One can find $\theta_{0}>0$ such that for $\vert\theta\vert<\theta_{0}$: 
\begin{enumerate}
\item This power series is absolutely uniformly convergent in $n$ and $\theta$. 
\item The following bound holds 
\begin{equation}
\bigl\vert\frac{1}{n}\ln Z_{{\rm small}}\bigr\vert\leq\frac{4}{n}z_{0}\sum_{x\in V\cup C}\sum_{{\gamma\ni x,{\rm \left\vert \gamma\right\vert <\lambda n}}}\theta^{\frac{c}{2}\vert\gamma\vert}e^{\left\vert \gamma\right\vert }.\label{eq:eq:polymer_inequality}
\end{equation}

\end{enumerate}
\end{lem}
\begin{IEEEproof}
When $\Gamma$ is an expander we can use the bound \eqref{eq:bound-activity}
on the activities of the small polymers. The proof is then almost
identical to that of Lemma \ref{th:lemmahightemp}. 
\end{IEEEproof}
Lemma \ref{th:convergence-lemma} has the following consequence (we
now take $z=1$): 
\begin{cor}
\label{th:lemmepol} Suppose $r>2$. Let $E$ be the event that $\Gamma$
is $(\lambda,\kappa)$ expander. For $\vert\theta\vert<\theta_{0}$,
\begin{equation}
\mathbb{E}_{\Gamma}\biggl[\frac{1}{n}\bigl\vert\ln Z_{{\rm small}}\bigr\vert\biggl\vert E\biggr]=O(n^{-(1-\chi)})\label{eq:coro}
\end{equation}
for any $0<\chi<1$. \end{cor}
\begin{rem}
We stress that Corollary \ref{th:lemmepol} and Lemma \ref{th:convergence-lemma}
hold for any $(l,r)$ with $r>2$. The restriction to odd $l$ will
come only when we estimate the contribution of large polymers. \end{rem}
\begin{IEEEproof}
Taking the conditional expectation over expander graphs \eqref{eq:eq:polymer_inequality}
implies 
\begin{eqnarray}
\frac{1}{n}\mathbb{E}_{\Gamma}\biggl[\bigl\vert\ln Z_{{\rm small}}\bigr\vert\biggl\vert E\biggr] & \leq & \mathbb{E}_{\Gamma}\biggl[\sum_{{\gamma\ni o,{\rm \left\vert \gamma\right\vert <\lambda n}}}\theta^{\frac{c}{2}\vert\gamma\vert}e^{\left\vert \gamma\right\vert }\biggl\vert E\biggr]\nonumber \\
 &  & \times4z_{0}\left(1+\frac{l}{r}\right).\label{eq:expectation}
\end{eqnarray}
We can compute this expectation by conditioning on the first event
that $\Gamma$ is tree-like in a neighborhood of size $O(\ln n)$
around this vertex, and on the second complementary event. The second
event has small probability $O(n^{-(1-\chi)})$ for any $0<\chi<1$.
Besides, from \eqref{eq:eq:polymer_inequality} it is easy to show
that $n^{-1}\vert\ln Z_{{\rm small}}\vert$ is bounded uniformly in
$n$. Thus the second event contributes only $O(n^{-(1-\chi)})$ to
the expectation. For the first event we have that the smallest polymer
is a cycle with $\vert\gamma\vert=O(\ln n)$. This implies that this
event contributes $O((\theta^{\frac{c}{2}}e^{A+1})^{\ln n}))$ to
the expectation. For small $\theta$ it is $O(n^{-(1-\chi)})$ that
dominates. 
\end{IEEEproof}

\subsection{Probability Estimates on Graphs\label{sec:probability on graphs}}

The loop sum is equal to the partition function of the gas of small
polymers plus a contribution containing at least one polymer of large
size $\vert\gamma\vert>\lambda n$. We call the later contribution
$R_{{\rm large}}$. More precisely 
\begin{equation}
1+\sum_{g\subset\Gamma}K(g)=Z_{{\rm small}}+R_{{\rm large}},
\end{equation}
where 
\begin{equation}
R_{{\rm large}}=\sum_{g\subset\Gamma{~{\rm s.t~}}\exists\gamma\subset g{~{\rm with~}}\vert\gamma\vert\geq\lambda n}K(g),
\end{equation}
The next lemma shows that the contribution from large polymers is
exponentially small, with high probability with respect to the graph
ensemble. 
\begin{lem}
\label{th:McK lemma} Fix $\delta>0$. Assume $l\geq3$ odd and $l<r$.
There exists a constant $C>0$ depending only on $l$ and $r$ such
that for $\theta$ small enough 
\begin{equation}
\mathbb{P}_{\Gamma}\biggl[\vert R_{{\rm large}}\vert\geq\delta\biggr]\leq\frac{1}{\delta}e^{-Cn}.\label{eq:proba_large_loop}
\end{equation}

\end{lem}
The proof which relies on counting estimates for subgraphs is presented
in the Appendix \ref{app:MCKay}. Unfortunately it breaks down for
$l$ even.

\subsection{Proof of Theorem \ref{th:theorem1}\label{sec:End of the proof}}

The results of Sections \ref{subsec:small polymer contribution} and
\ref{sec:probability on graphs} allow us to prove the following. 
\begin{prop}
\label{th:second-theorem} Suppose $l$ is odd and $3\leq l<r$. Take
$\Gamma$ at random in $\mathcal{B}(l,r,n)$. There exist a small
$\theta_{0}$ independent of $n$ such that for $\theta<\theta_{0}$,
and any high-noise solution $\left(\underline{\eta},\underline{\widehat{\eta}}\right)$
of the BP equations, with probability $1-O(n^{-(l(1-\kappa)-1)})$
we have 
\begin{align}
\bigl\vert\frac{1}{n}\ln Z_{{\rm LDPC}}-(f_{{\rm LDPC}}^{\mathrm{Bethe}}\left(\underline{\eta},\underline{\widehat{\eta}}\right)+ & \frac{1}{n}\ln Z_{{\rm small}})\bigr\vert\nonumber \\
 & =O(e^{-n^{l(1-\kappa)-1}})\,.\label{eq:second-result}
\end{align}
\end{prop}
\begin{rem}
We recall that $0<l(1-\kappa)-1<(r-2)/r$. This proposition shows
that large polymers contribute only with exponentially small corrections
to the Bethe free energy. Inverse power in $n$ corrections can be
computed systematically from the polymer expansion of $n^{-1}\ln Z_{{\rm small}}$. \end{rem}
\begin{IEEEproof}
Note that 
\begin{equation}
\frac{1}{n}\ln\biggl\{\sum_{g\subset\Gamma}K(g)\biggr\}=\frac{1}{n}\ln Z_{{\rm small}}+\frac{1}{n}\ln\biggl(1+\frac{R_{{\rm large}}}{Z_{{\rm small}}}\biggr),
\end{equation}
which means that the term on the left hand side of \eqref{eq:second-result}
is equal to 
\begin{equation}
\frac{1}{n}\biggl\vert\ln\biggl(1+\frac{R_{{\rm large}}}{Z_{{\rm small}}}\biggr)\biggr\vert\,.
\end{equation}

On one hand, from Corollary \ref{th:lemmepol} and the Markov bound,
we have for any $\epsilon>0$, 
\begin{equation}
\mathbb{P}\biggl[e^{-n\epsilon}\leq Z_{{\rm small}}^{-1}\leq e^{n\epsilon}\biggl\vert E\biggr]=1-\frac{1}{\epsilon}O(n^{-(1-\chi)}).\label{eq:strange}
\end{equation}
On the other hand, from Lemma \ref{th:McK lemma} 
\begin{align}
\mathbb{P}\biggl[\left|R_{{\rm large}}\right|\geq\delta\biggl\vert E\biggl]\mathbb{P}[E]\leq\mathbb{P}\biggl[\left|R_{{\rm large}}\right|\geq\delta\biggl]\leq\frac{1}{\delta}e^{-Cn}
\end{align}
and since $\mathbb{P}[E]=1-O(n^{-(l(1-\kappa)-1)})$, 
\begin{equation}
\mathbb{P}\biggl[\left|R_{{\rm large}}\right|\geq\delta\biggl\vert E\biggl]\leq\frac{2}{\delta}e^{-Cn}.\label{eq:variant}
\end{equation}
Using \eqref{eq:strange} and \eqref{eq:variant}, and choosing $\frac{\delta}{2}=e^{-2n\epsilon}$
it is not difficult to show that (at this point one must take $0<2\epsilon<C$)
\begin{equation}
\mathbb{P}\biggl[\biggl\vert\frac{R_{{\rm large}}}{Z_{{\rm small}}}\biggr\vert\geq e^{-n\epsilon}\biggl\vert E\biggr]\leq\frac{1}{\epsilon}O(n^{-(1-\chi)})+e^{-n(C-2\epsilon)}\,.
\end{equation}
Thus 
\begin{align}
\mathbb{P}\biggl[ & \biggl\vert\frac{R_{{\rm large}}}{Z_{{\rm small}}}\biggr\vert\leq e^{-n\epsilon}\biggl\vert E\biggr]\geq1-\frac{1}{\epsilon}O(n^{-(1-\chi)})\,.
\end{align}
This implies for $n$ large 
\begin{align}
\mathbb{P}\biggl[ & \biggl\vert\ln\biggl(1+\frac{R_{{\rm large}}}{Z_{{\rm small}}}\biggr)\biggr\vert\leq2e^{-n\epsilon}\biggl\vert E\biggr]\geq1-\frac{1}{\epsilon}O(n^{-(1-\chi)}).\label{eq:88}
\end{align}
Furthermore 
\begin{align}
\mathbb{P}\biggl[ & \biggl\vert\ln\biggl(1+\frac{R_{{\rm large}}}{Z_{{\rm small}}}\biggr)\biggr\vert\leq2e^{-n\epsilon}\biggr]\nonumber \\
 & \geq\bigr(1-\frac{1}{\epsilon}O(n^{-(1-\chi)})\bigl)\bigl(1-O(n^{-(l(1-\kappa)-1)})\bigr)\nonumber \\
 & \geq1-O(n^{-(l(1-\kappa)-1)})\,.\label{eq:atlast}
\end{align}
The last line is obtained by choosing 
\begin{equation}
\epsilon=\frac{n^{l(1-\kappa)-1}}{n^{1-\chi}}\leq n^{\frac{r-2}{r}-1+\chi}<\frac{C}{2},\label{eq:choose-epsilon}
\end{equation}
which is possible since $\kappa\in]1-\frac{2(r-1)}{lr},1-\frac{1}{l}[$
and we can take $\chi>0$ as small as we wish.

Finally from 
\begin{equation}
n\epsilon=n^{l(1-\kappa)-1+\chi}\geq n^{l(1-\kappa)-1}
\end{equation}
and \eqref{eq:atlast} we deduce the statement of the proposition. 
\end{IEEEproof}
It is now possible to complete the proof of Theorem \ref{th:theorem1}.
\begin{IEEEproof}[Proof of Theorem \ref{th:theorem1}]
 Consider the difference 
\begin{equation}
\bigl\vert\frac{1}{n}\ln Z_{{\rm LDPC}}-f_{{\rm LDPC}}^{{\rm Bethe}}(\underline{\eta},\widehat{\underline{\eta}})\bigr\vert\label{eq:finaldifference}
\end{equation}
We first remark that this quantity is bounded uniformly in $n$ because
each term $n^{-1}\vert\ln Z_{{\rm LDPC}}\vert$ and $\vert f_{{\rm LDPC}}^{{\rm Bethe}}\vert$
is bounded, as can be checked directly from their definition.

Now consider the event $S$ - or the set of graphs - such that 
\begin{equation}
\biggl\vert\ln\biggl(1+\frac{R_{{\rm large}}}{Z_{{\rm small}}}\biggr)\biggr\vert\leq e^{-n^{l(1-\kappa)-1}}.
\end{equation}
Proposition \eqref{th:second-theorem} says that 
\begin{equation}
\mathbb{P}[S^{c}]=O(n^{-(l(1-\kappa)-1)}).
\end{equation}
Thus we have 
\begin{equation}
\mathbb{E}\biggl[\bigl\vert\frac{1}{n}\ln Z_{{\rm LDPC}}-f_{{\rm LDPC}}^{{\rm Bethe}}\bigr\vert\biggl\vert S^{c}\biggr]\mathbb{P}[S^{c}]=O(n^{-(l(1-\kappa)-1)})\,.\label{eq:95}
\end{equation}

We will now estimate 
\begin{equation}
\mathbb{E}\biggl[\bigl\vert\frac{1}{n}\ln Z_{{\rm LDPC}}-f_{{\rm LDPC}}^{{\rm Bethe}}\bigr\vert\biggl\vert S\biggr]\mathbb{P}[S]\,.
\end{equation}
Since $\mathbb{P}[S]=1-O(n^{-(l(1-\kappa)-1)})$ we have to show that
the expectation conditioned over $S$ is small. 
\begin{align}
\mathbb{E}\biggl[\bigl\vert\frac{1}{n}\ln Z_{{\rm LDPC}} & -f_{{\rm LDPC}}^{{\rm Bethe}}\bigr\vert\biggl\vert S\biggr]\nonumber \\
 & =\mathbb{E}\biggl[\bigl\vert\frac{1}{n}\ln Z_{{\rm LDPC}}-f_{{\rm LDPC}}^{{\rm Bethe}}\bigr\vert\biggl\vert S\cap E\biggr]\mathbb{P}[E]\nonumber \\
 & +\mathbb{E}\biggl[\bigl\vert\frac{1}{n}\ln Z_{{\rm LDPC}}-f_{{\rm LDPC}}^{{\rm Bethe}}\bigr\vert\biggl\vert S\cap E^{c}\biggr]\mathbb{P}[E^{c}].\label{eq:86}
\end{align}
Since, as remarked before, \eqref{eq:finaldifference} is bounded
and $\mathbb{P}[E^{c}]=O(n^{-(l(1-\kappa)-1})$ the second term on
the right hand side is $O(n^{-(l(1-\kappa)-1})$. It remains to show
that the conditional expectation in the first term of the right hand
side is small. This is bounded above by two contributions. The first
one is 
\begin{align}
\mathbb{E}\biggl[\frac{1}{n}\biggl\vert\ln\biggl(1+\frac{R_{{\rm large}}}{Z_{{\rm small}}}\biggr)\biggr\vert\biggl\vert S\cap E\biggr]\leq e^{-n^{l(1-\kappa)-1}}\,,\label{eq:98}
\end{align}
and the second (recall Corollary \ref{th:lemmepol}) 
\begin{equation}
\mathbb{E}\biggl[\frac{1}{n}\vert\ln Z_{{\rm small}}\vert\biggl\vert S\cap E\biggr]=O(n^{-(1-\chi)})\,.\label{eq:99}
\end{equation}

Putting all contributions \eqref{eq:95}, \eqref{eq:86}, \eqref{eq:98},
\eqref{eq:99} together we obtain the desired result 
\begin{align}
\mathbb{E}\biggl[\bigl\vert\frac{1}{n}\ln Z_{{\rm LDPC}}-f_{{\rm LDPC}}^{{\rm Bethe}}\bigr\vert\biggr] & =O(n^{-(1-\chi)})+O(n^{-(l(1-\kappa)-1)})\nonumber \\
 & =O(n^{-(l(1-\kappa)-1)})\,.
\end{align}
In the last step we have taken $0<\chi<l(1-\kappa)-1$. 
\end{IEEEproof}

\section{Discussion\label{sec:discussion}}

\subsection{LDPC: The Case $l$ Even}

When $l$ is even the point $\theta=0$ as a singular behavior. As
the channel realization is trivial $\underline{h}=\underline{0}$,
the low-noise fixed point is simply the all zeros messages $\left(\eta_{i\rightarrow a},\widehat{\eta}_{a\rightarrow i}\right)=\left(0,0\right)$.
The activities can be computed exactly for this BP fixed point
\begin{equation}
K_{a}\left(\gamma\right)=\begin{cases}
1 & {\rm if}\left|\partial a\cap\gamma\right|=r\\
0 & {\rm otherwise}
\end{cases},\,K_{i}\left(\gamma\right)=\frac{1+\left(-1\right)^{\left|\partial i\cap\gamma\right|}}{2}.
\end{equation}
When the graph $\Gamma$ is an expander, every small polymer $\left|\gamma\right|<n\lambda$
contains at least one check node with induced degree less than $r$
(see Lemma \ref{th:expander_activity}). Thus $K\left(\gamma\right)=0$
and 
\begin{equation}
Z_{{\rm small}}=1.
\end{equation}
The contribution of the small polymer vanishes which is of course
in adequacy with the prediction of the polymer expansion (see Lemma
\ref{th:convergence-lemma}). 

However for the total graph, and unlike the case $l$ odd, we have
\begin{equation}
K\left(\Gamma\right)=1.
\end{equation}
More generally polymers with a size of the order of the total graph
have an activity close to one. This implies that the contribution
coming from large polymer is non-vanishing but is growing linearly
\begin{equation}
1<\mathbb{E}_{\Gamma}\left(\left|R_{{\rm large}}\right|\right)<C_{l,r}n^{4r^{2}},
\end{equation}
as it can be shown using the same counting arguments as in the Appendix
\ref{app:MCKay}. As a consequence, we find similarly to Theorem \ref{th:second-theorem}
that the Bethe free energy is asymptotically exact with high probability.
More precisely, with probability $1-O(n^{-(l(1-\kappa)-1)})$ 
\begin{equation}
\bigl\vert\frac{1}{n}\ln Z_{{\rm LDPC}}-f_{{\rm LDPC}}^{{\rm Bethe}}\bigr\vert=O\left(\frac{1}{n}\ln n\right).
\end{equation}
The notable difference with Theorem \ref{th:second-theorem} is that
the decay rate of the difference is not exponential.

When $l$ is even and $\theta>0$, the bound on the activity of the
total graph predict an exponential growth
\begin{equation}
\overline{K}\left(\Gamma\right)=\left(1+\alpha_{1}\theta^{r}\right)^{\frac{l}{r}n}\left(1+\alpha_{2}\theta^{2}\right)^{n}.
\end{equation}
The contribution of the large polymer can no longer be estimated as
in Appendix \ref{app:MCKay}. To tackle this problem, it seems necessary
to have a precise control of sign cancellations in the sum $R_{{\rm large}}$.
Such a control is out of reach of the method presented in the paper.

\subsection{The Case $l>r$}

The constraint $r>l$ appears naturally for proving Theorem \ref{th:McK lemma}.
If $r<l$, large polymer with an activity exponentially increasing
with their size are also exponentially numerous. Therefore we found
using counting arguments that the contribution of $R_{{\rm large}}$
is not negligible.

When $r<l$, the graphical model is no longer describing a code. In
fact for at exactly $h=0$, the partition function counts the number
of solution of a random linear system of equations which is overdetermined.
This corresponds to an UNSAT phase of a linear constraints satisfaction
problem. In this phase, it is expected that the Bethe free energy
is not a good approximation of the free energy (instead one should
use the RSB free energy \cite{mezard09information}). It seems then
reasonable to think that the corrections to the Bethe free energy
are non vanishing even with a precise control of the sign cancellation
in the activities.

\subsection{LDPC Low Noise}

The low noise regime is characterized by half-log likelihoods with
high magnitudes $h\approx\infty$. The low noise fixed point of the
Belief-Propagation equations is the trivial solution
\begin{equation}
\left(\eta_{i\rightarrow a},\widehat{\eta}_{a\rightarrow i}\right)=\left(+\infty,+\infty\right).
\end{equation}
The activities can be computed exactly at the low noise fixed point
\begin{eqnarray}
K\left(\gamma\right) & = & \prod_{a\in\gamma\cap C}\frac{1+\left(-1\right)^{\left|\partial a\cap\gamma\right|}}{2}\nonumber \\
 & \times & \prod_{i\in\gamma\cap V}\left(-1\right)^{l}e^{-2h_{i}}\mathbb{I}\left(\left|\partial i\cap\gamma\right|=l\right).\label{eq:low noise activity}
\end{eqnarray}
According to \eqref{eq:low noise activity}, polymers are subgraphs
which have check nodes with even induced degree and variable nodes
with induced degree equal to $l$. The particularity of the low noise
activities is that their intensity depends on the sign of the half-log
likelihoods$h_{i}$, and a fortiori on the distribution of $\underline{h}$.
Using Hoeffding's inequality, we see that a polymer have, with large
probability, a small activity 
\begin{equation}
\mathbb{P}_{\underline{h}}\left(\left|K\left(\gamma\right)\right|\leq e^{-2h\left(\tanh h-2\epsilon\right)\left|\gamma\cap V\right|}\right)\geq1-e^{-2\epsilon^{2}\left|\gamma\cap V\right|}.
\end{equation}
However, the expected activity is dominated by rare events
\begin{equation}
\mathbb{E}_{\underline{h}}\left[\left|K\left(\gamma\right)\right|\right]=1.\label{eq:low noise expected activity}
\end{equation}
This prevents to use the same counting arguments as in Appendix \ref{app:MCKay}.

\subsection{Lattice Graph with Low Dimension}

\begin{figure}[tbh]

\centering{}\includegraphics[scale=0.1]{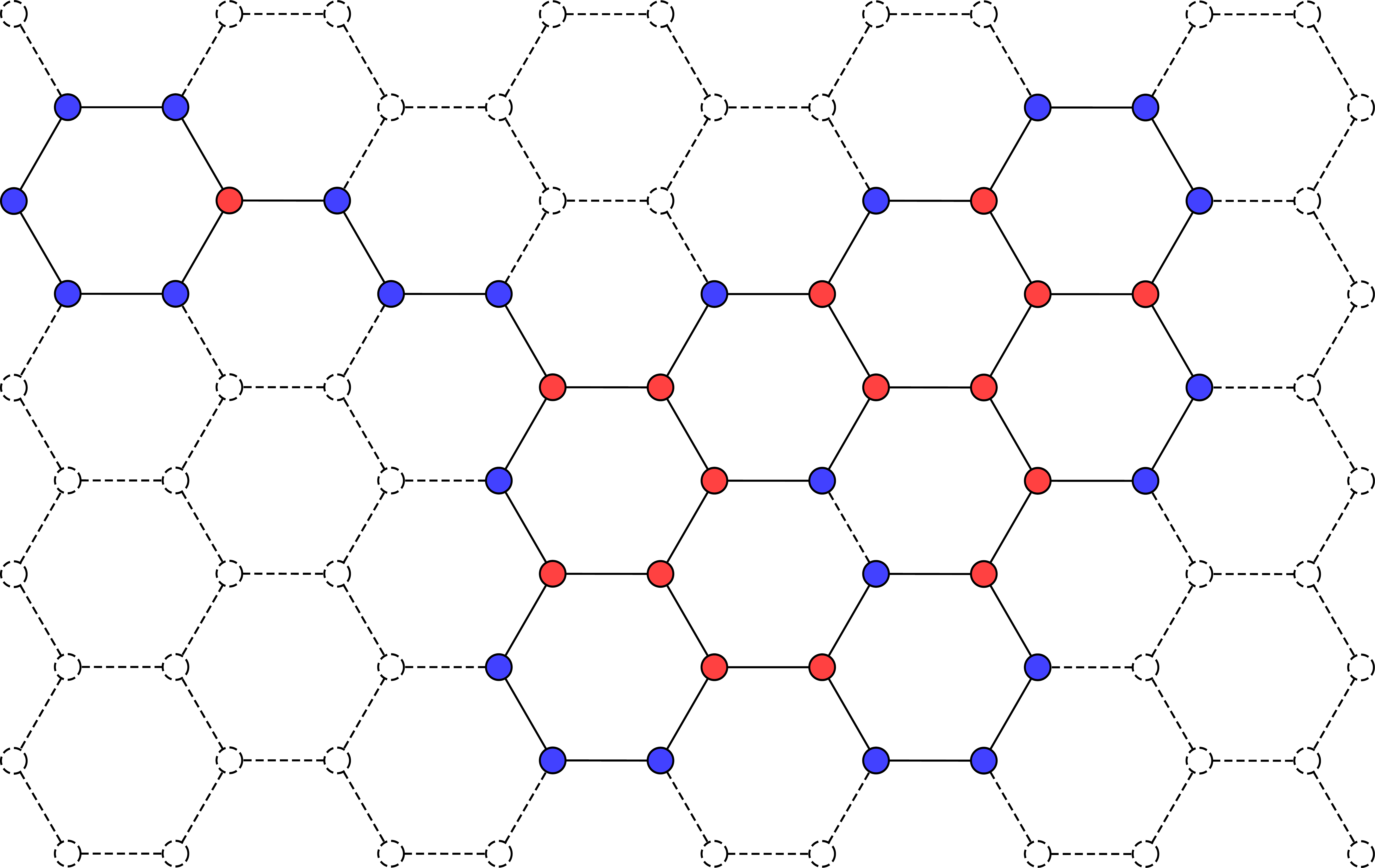}\protect\caption{\label{fig:honeycom}Polymer in a honeycomb lattice at low temperature.
The activity is decreasing in the boundary size (blue nodes).}
\end{figure}

An other possible application of the polymer expansion is the study
of spin systems on a lattice $\Lambda=\left(V,E\right)$. In low dimension,
it is known that the mean field approximation like the Bethe free
energy is equal to the free energy and thus 
\begin{equation}
\lim_{n\rightarrow\infty}\frac{1}{n}\ln Z_{{\rm polymer}}\neq0.
\end{equation}
However, if the polymer expansion converges, it could be use as a
systematic way of computing corrections to the mean field approximation.

Let us illustrated the question of the convergence for the Ising model
on the honeycomb lattice (see Figure \ref{fig:honeycom}). The spins
are attached to vertices and an edge represents a ferromagnetic interaction
between spins
\begin{equation}
Z_{{\rm Ising}}=\sum_{\underline{\sigma}\in\left\{ -1,1\right\} ^{n}}\prod_{\left(i,j\right)\in E}\exp\left(\beta J\sigma_{i}\sigma_{j}\right).
\end{equation}
There are three solutions of the BP equations and they can be computed
exactly. There is one fixed point which described a high-temperature
phase ($\beta J\approx0$) and two fixed point describing a low-temperature
phase ($\beta J\approx\infty$). The two low-temperature fixed point
differs only by a sign.

The high-temperature activity is

\begin{equation}
K_{{\rm Ising}}^{0}\left(\gamma\right)=\begin{cases}
\tanh\beta J & {\rm if}\,\gamma\,{\rm is\,a\,cycle}\\
0 & {\rm otherwise}
\end{cases}.
\end{equation}
Thus the activity is decreasing in the polymer size. To prove the
convergence of the expansion for high temperature one can apply directly
Lemma \ref{th:lemmahightemp}. Or one can see that, in the honeycomb
lattice, the number of cycles containing the same vertex and having
a length $t$ is upper-bounded by $2^{t}.$ 

For small temperature the activity is

\begin{eqnarray}
K_{{\rm Ising}}^{\pm\infty}\left(\gamma\right) & = & \prod_{i\in V,\left|\partial i\cap\gamma\right|=3}\frac{1+\tanh\beta J}{2\tanh\beta J}\sqrt{\frac{2\tanh\beta J-1}{\tanh\beta J}}\nonumber \\
 & \times & \prod_{i\in V,\left|\partial i\cap\gamma\right|=2}\mp\frac{1-\tanh\beta J}{2\tanh\beta J}.
\end{eqnarray}
The polymers have an activity which is decreasing with respect to
the size of the boundary (the nodes with induced degree equal to two).
This is similar to the activity of polymers for LDPC at high noise.
But unlike the LDPC case, we cannot apply an expander argument to
prove that the boundary of a polymer is of the same order than its
size. In fact we suspect that the polymer series is not convergent (at low temperatures) but that
it is an asymptotic series. This is supported by the fact that for exactly solvable models (e.g. Ising 
in two dimensions and Ice model)
the first few loop corrections allow
to compute refined approximations to the true free energy. We hope to come back to this question elsewhere.


\appendices{}

\section{Derivation of the Loop-Sum Identity\label{app:loop sum identity}}

The ``loop-sum identity'' is a representation of the error term
between the free energy and the Bethe free energy. It takes the form
of the logarithm of a sum over sub-graphs that are non-necessarily
connected. This identity was first derived for graphical models with
binary variables by Chertkov and Chernyak in \cite{chertkov2006loop}
and later generalized for variables on a $q$-ary alphabet by the
same authors in \cite{chernyak2007loop}. The extension of the loop-sum
identity to continuous alphabet has been carried out by Xiao and Zhou
in \cite{Xiao2011}. The present section contains a short derivation
of the loop-sum identity based on the original paper \cite{chertkov2006loop}.
There exists other representations of the loop-sum identity and its
generalization notably as the holographic transformation of a normal
factor graph \cite{G.D.Forney2011,mori2013thesis}.

Consider the problem of computing the partition function of a factor
graph model 
\begin{equation}
Z=\sum_{\underline{s}\in\{-1,+1\}^{n}}\prod_{a\in C}\psi_{a}(s_{\partial a}).\label{eq:spin_system}
\end{equation}
The loop expansion takes a natural form on graphical models called
vertex models, where variables are attached to edges. We introduce
the auxiliary set of spins $\sigma_{ia},\widehat{\sigma}_{ai}\in\{-1,1\}$
attached to directed edges $(i\rightarrow a)$ and $(a\rightarrow i)$
respectively. We denote by $\sigma_{\partial a}=\{\sigma_{ja}\vert j\in\partial a\}$
the collection of spins that are on edges pointing toward $a$ and
we denote by $\widehat{\sigma}_{\partial i}=\{\widehat{\sigma}_{bi}\vert b\in\partial i\}$
the collection of spins that are on edges pointing toward $i$. We
can rewrite \eqref{eq:spin_system} as a partition function of a vertex
model 
\begin{equation}
Z=\sum_{\underline{\sigma},\underline{\widehat{\sigma}}\in\{-1,1\}^{\vert E\vert}}\prod_{a\in C}\psi_{a}(\sigma_{\partial a})\prod_{i\in V}\phi_{i}(\widehat{\sigma}_{\partial i})\prod_{(a,i)\in E}\frac{1+\sigma_{ia}\widehat{\sigma}_{ai}}{2},\label{eq:spin_system_vertex}
\end{equation}
where 
\begin{equation}
\phi_{i}(\widehat{\sigma}_{\partial i})=\prod_{b,c\in\partial i}\frac{1+\widehat{\sigma}_{bi}\widehat{\sigma}_{ci}}{2}.
\end{equation}

Let us comment on the expression \eqref{eq:spin_system_vertex}. The
new factors $\phi_{i}(\widehat{\sigma}_{\partial i})$ ensure that
all spins on edges outgoing from a variable node $i$ take the same
value $s_{i}$. As for the last product, it forces spins on the same
edge to be equal. The key idea in the loop expansion is to ``soften''
the constraints on the edges before performing the expansion.. Using
the following identity, valid for any binary distributions $\nu_{ia}$
and $\widehat{\nu}_{ai}$ ,

\begin{equation}
\frac{1+\sigma_{ia}\widehat{\sigma}_{ai}}{2}=\frac{\nu_{ia}(\sigma_{ia})\widehat{\nu}_{ai}(\widehat{\sigma}_{ai})+\sigma_{ia}\widehat{\nu}_{ai}(-\sigma_{ia})\widehat{\sigma}_{ai}\nu_{ia}(-\widehat{\sigma}_{ai})}{\sum_{s\in\{-1,1\}}\nu_{ia}(s)\widehat{\nu}_{ai}(s)},\label{eq:delta_softening}
\end{equation}

we can rewrite the partition function \eqref{eq:spin_system_vertex}
\begin{align}
Z=\sum_{\underline{\sigma},\underline{\widehat{\sigma}}\in\{-1,1\}^{\vert E\vert}} & \prod_{a\in C}\psi_{a}(\sigma_{\partial a})\prod_{j\in\partial{a}}\nu_{ja}(\sigma_{ja})\nonumber \\
 & \prod_{i\in V}\phi_{i}(\widehat{\sigma}_{\partial i})\prod_{b\in\partial{i}}\widehat{\nu}_{bi}(\widehat{\sigma}_{bi})\nonumber \\
 & \prod_{(a,i)\in E}\left(\sum_{s\in\{-1,1\}}\nu_{ia}(s)\widehat{\nu}_{ai}(s)\right)^{-1}\nonumber \\
 & \prod_{(a,i)\in E}\left(1+\sigma_{ia}\frac{\widehat{\nu}_{ai}(-\sigma_{ia})}{\nu_{ia}(\sigma_{ia})}\widehat{\sigma}_{ai}\frac{\nu_{ia}(-\widehat{\sigma}_{ai})}{\widehat{\nu}_{ai}(\widehat{\sigma}_{ai})}\right).\label{eq:z_before_expansion}
\end{align}

We use the ``generalized binomial formula'' on graphs. For any function
$f$ defined on the edges $e\in E$ of a graph $\Gamma$, the following
relation holds 
\begin{equation}
\prod_{e\in E}(1+f(e))=1+\sum_{g\subset\Gamma}\prod_{e\in E\cap g}f(e),\label{eq:binomial_theorem}
\end{equation}
where the sum runs on every non-empty subset of edges represented
by subgraphs $g$ whose vertices are incident to the edges in the
subset. On the left-hand side of Equation \eqref{eq:binomial_theorem}
the products run over the set of edges of $g$ which is denoted by
$E\cap g$. Expanding the last product in \eqref{eq:z_before_expansion}
with the generalized binomial formula leads to

\begin{align}
Z=\exp(nf^{\mathrm{Bethe}})\times\biggl(1+\sum_{g\subset\Gamma}K\left(g\right)\biggr).\label{eq:loop_sum_identity}
\end{align}
The quantity that factorized in the expansion appears to be the Bethe
free energy 
\begin{equation}
f^{\mathrm{Bethe}}(\underline{\nu},\underline{\widehat{\nu}})=\frac{1}{n}\left(\sum_{a\in C}F_{a}+\sum_{i\in V}F_{i}-\sum_{(i,a)\in E}F_{ia}\right),
\end{equation}

where

\begin{align}
\begin{cases}
F_{a}=\ln\{\sum_{s_{\partial a}}\psi_{a}(s_{\partial a})\prod_{j\in\partial a}\nu_{ja}(s_{j})\},\\
\\
F_{i}=\ln\{\sum_{s_{i}}\prod_{b\in\partial i}\widehat{\nu}_{bi}(s_{i})\},\\
\\
F_{ia}=\ln\{\sum_{s_{i}}\nu_{ia}(s_{i})\widehat{\nu}_{ai}(s_{i})\}.
\end{cases}.
\end{align}
The activities $K(g)$ associated with each subgraphs can be distributed
in contributions coming from vertices in $g$ 
\begin{align}
K(g)=\prod_{i\in g\cap V}K_{i}\prod_{a\in g\cap C}K_{a},\label{eq:total_activity}
\end{align}
where 
\begin{align}
K_{i}(g)=\frac{\sum_{s_{i}}\prod_{a\in\partial i\setminus g}\widehat{\nu}_{ai}(s_{i})\prod_{a\in\partial i\cap g}s_{i}\nu_{ia}(-s_{i})}{\sum_{s_{i}}\prod_{a\in\partial i}\widehat{\nu}_{ai}(s_{i})},\label{eq:Ki}
\end{align}
and 
\begin{align}
K_{a}(g)=\frac{\sum_{s_{\partial a}}\psi_{a}(s_{\partial{a}})\prod_{i\in\partial a\setminus g}\nu_{ia}(s_{i})\prod_{i\in\partial a\cap g}s_{i}\widehat{\nu}_{ai}(-s_{i})}{\sum_{s_{\partial a}}\psi_{a}(s_{\partial{a}})\prod_{i\in\partial a}\nu_{ia}(s_{i})}.\label{eq:Ka}
\end{align}
The sum over subgraphs in \eqref{eq:loop_sum_identity} is the ``loop-sum
identity''. Note that for the moment the binary distributions entering
in \eqref{eq:delta_softening} are completely arbitrary. The transformation
\eqref{eq:delta_softening} is crucial in that it allows the preservation
of the correlations between neighboring spins. Messages $\widehat{\nu}_{a\rightarrow i}$
directed toward a spin $\sigma_{i}$ can be interpreted as an interaction
from the neighboring variables $\underline{\sigma}_{\partial a\setminus i}$
that bias the average value of the spin $\sigma_{i}$. Expanding the
Kronecker delta in \eqref{eq:spin_system_vertex} directly is equivalent
as taking $\underline{\nu}$ and $\underline{\widehat{\nu}}$ as being
uniform distributions. Such an expansion would be accurate only in
a regime were the spins are almost independent from each other and
almost uniformly distributed between $+1$ and $-1$. This is the
case for instance in the high-temperature regime. Thanks to the transformation
\eqref{eq:delta_softening}, the effect of correlations between neighboring
spins can be captured by the distributions $\underline{\nu}$ and
$\underline{\widehat{\nu}}$. Thus, for appropriate choices of messages
$\left(\underline{\nu},\widehat{\underline{\nu}}\right)$, the expansion
can also be accurate in the low-temperature regime.

In order for the loop-sum identity to be useful one has to choose
the ``correct'' binary distributions. The natural requirement for
sparse locally tree-like graphs is that every subgraph $g$ that is
not a loop must have a zero weight. Said differently, the distributions
$\nu$ and $\widehat{\nu}$ must be chosen such that $\vert\partial a\cap g\vert=1$
and $\vert\partial i\cap g\vert=1$ implies $K_{a}(g)=0$ and $K_{i}(g)=0$
respectively. The requirement $K_{a}(g)=0$ is fulfilled by distributions
$\widehat{\nu}_{ai}$ that satisfy the following equation 
\begin{align}
\sum_{s_{i}}s_{i}\widehat{\nu}_{ai}(-s_{i})\sum_{s_{\partial a\setminus i}}\psi_{a}(s_{\partial a})\prod_{j\in\partial a\setminus i}\nu_{ja}(s_{j})=0.
\end{align}
This is satisfied if $\widehat{\nu}_{ai}$ is a solution of the first
Belief-Propagation equation 
\begin{align}
\widehat{\nu}_{ai}(s_{i})=\frac{\sum_{s_{\partial a\setminus i}}\psi_{a}(s_{\partial a})\prod_{j\in\partial a\setminus i}\nu_{ja}(s_{j})}{\sum_{s_{\partial a}}\psi_{a}(s_{\partial a})\prod_{j\in\partial a\setminus i}\nu_{ja}(s_{j})}.\label{eq:BP_equation_ai}
\end{align}
Similarly one can check that the requirement $K_{i}(g)=0$ is fulfilled
by the choice 
\begin{align}
\nu_{ia}(s_{i})=\frac{\prod_{b\in\partial i\setminus a}\widehat{\nu}_{bi}(s_{i})}{\sum_{s_{i}}\prod_{b\in\partial i\setminus a}\widehat{\nu}_{bi}(s_{i})}.\label{eq:BP_equation_ia}
\end{align}
This is nothing else but the second belief propagation equation.

\section{Polymer-Expansion Identity \eqref{eq:polymerexpansion}\label{app:polymer} }

The polymer expansion is a powerful tool from statistical physics
to expand the logarithm of a polymer partition function in powers
of the activity. We give in this appendix a quick derivation of the
polymer expansion based on \cite{brydges1984short}.

We recall that $\mathcal{G}_{M}$ is the set of all connected graphs
$G$ with $M$ labeled vertices $1,\ldots,M$ and that the Ursell
functions are
\begin{equation}
U_{M}(\gamma_{1},\ldots,\gamma_{M})=\sum_{G\in\mathcal{G}_{M}}\prod_{(k,k^{\prime})\in G}\left(\mathbb{I}\left(\gamma_{k}\cap\gamma_{k^{\prime}}=\emptyset\right)-1\right),\label{eq:appUrsell}
\end{equation}
if $M>1$ and $U_{1}\left(\gamma\right)\equiv1$ otherwise. We recall
that the notation $\prod_{(k,k^{\prime})\in G}$ in Equation \eqref{eq:appUrsell}
is a shorthand for the product over the edges $\left(k,k'\right)$
of $G$. Furthermore we denote the complete graph with $M$ labeled
vertices by $\mathcal{K}_{M}$. We say that a partition of the set
$\left\{ 1,\ldots,M\right\} $ into $q$ ``blocks'' is an unordered
list $\left\{ \mathcal{I}_{1},\ldots,\mathcal{I}_{q}\right\} $ of
disjoint nonempty subsets $\mathcal{I}_{t}\subset\left\{ 1,\ldots,M\right\} $.
The partitions of $M$ elements into $q$ ``blocks'' form an ensemble
denoted by $\mathcal{P}_{M}^{q}$.

The polymer partition function is

\begin{align}
Z^{{\rm polymer}}=1+\sum_{M\geq1} & \frac{1}{M!}\sum_{\gamma_{1},...,\gamma_{M}\subset\Gamma}\prod_{k=1}^{M}K\left(\gamma_{k}\right)\nonumber \\
 & \times\prod_{k<k^{\prime}}\mathbb{I}\left(\gamma_{k}\cap\gamma_{k^{\prime}}=\emptyset\right).\label{eq:Appendix polymer partition function}
\end{align}
We recall that polymers $\gamma$ are connected subgraphs of $\Gamma$
that cannot intersect due to the presence of the hard core constraints
$\mathbb{I}\left(\gamma_{k}\cap\gamma_{k^{\prime}}=\emptyset\right)$.
The polymer-expansion identity is based on the expansion of these
hard core constraints using the binomial theorem on graphs (Eqn. \eqref{eq:binomial_theorem})

\begin{align}
\prod_{k<k^{\prime}} & \mathbb{I}\left(\gamma_{k}\cap\gamma_{k^{\prime}}=\emptyset\right)=\prod_{\left(k,k^{\prime}\right)\in\mathcal{K}_{M}}\left(\mathbb{I}\left(\gamma_{k}\cap\gamma_{k^{\prime}}=\emptyset\right)-1+1\right)\nonumber \\
 & =1+\sum_{G\subset\mathcal{K}_{M}}\prod_{(k,k^{\prime})\in G}\left(\mathbb{I}\left(\gamma_{k}\cap\gamma_{k^{\prime}}=\emptyset\right)-1\right),\label{eq:hard core expansion}
\end{align}
The sum in \eqref{eq:hard core expansion} runs over non-empty subset
of edges of $\mathcal{K}_{M}$ represented by subgraphs $G$ whose
vertices are incident to the edges in the subset. Notice that each
general subgraph in $\mathcal{K}_{M}$ can be written as an union
of disjoint connected subgraphs $G_{1},...,G_{q}$. This with the
fact that $U_{1}\left(\gamma_{k}\right)=1$ enable us to re-sum \eqref{eq:hard core expansion}
as 
\begin{eqnarray}
1 & + & \sum_{G\subset\mathcal{K}_{M}}\prod_{(k,k^{\prime})\in G}\left(\mathbb{I}\left(\gamma_{k}\cap\gamma_{k^{\prime}}=\emptyset\right)-1\right)\nonumber \\
 & = & \sum_{q=1}^{M}\sum_{\left\{ \mathcal{I}_{1},\ldots,\mathcal{I}_{q}\right\} \in\mathcal{P}_{M}^{q}}\prod_{t=1}^{q}U_{\left|\mathcal{I}_{t}\right|}\left(\left(\gamma_{k}\right)_{k\in\mathcal{I}_{t}}\right).\label{eq:partition over connected components}
\end{eqnarray}
Together with \eqref{eq:hard core expansion} and \eqref{eq:partition over connected components},
the polymer partition function \eqref{eq:Appendix polymer partition function}
can be rewritten as 
\begin{align}
Z^{{\rm polymer}}=1+\sum_{M\geq1} & \frac{1}{M!}\sum_{q=1}^{M}\sum_{\left\{ \mathcal{I}_{1},...,\mathcal{I}_{q}\right\} \in\mathcal{P}_{M}^{q}}\prod_{t=1}^{q}\phi\left(\mathcal{I}_{t}\right),\label{eq:polymer expansion phi}
\end{align}
where

\begin{equation}
\phi\left(\mathcal{I}_{t}\right):=\sum_{\gamma_{k\in\mathcal{I}_{t}}}U_{\left|\mathcal{I}_{t}\right|}\left(\left(\gamma_{k}\right)_{k\in\mathcal{I}_{t}}\right)\prod_{k\in\mathcal{I}_{t}}K\left(\gamma_{k}\right).\label{eq:phi It}
\end{equation}
The function introduced in \eqref{eq:phi It} depends only on the
size of the ensemble 
\begin{equation}
\phi\left(\mathcal{I}_{t}\right)=\phi\left(\left|\mathcal{I}_{t}\right|\right),
\end{equation}
as $k\in\mathcal{I}_{t}$ in \eqref{eq:phi It} are just dummy indices.
The number of partitions of $\left\{ 1,...,M\right\} $ with prescribed
size $\left|\mathcal{I}_{1}\right|=m_{1},...,\left|\mathcal{I}_{q}\right|=m_{q}$
is 
\begin{equation}
\sum_{\left\{ \left|\mathcal{I}_{1}\right|=m_{1},...,\left|\mathcal{I}_{q}\right|=m_{q}\right\} \in\mathcal{P}_{M}^{q}}1=\frac{M!}{q!}\prod_{t=1}^{q}\frac{1}{m_{t}!},
\end{equation}
where $m_{1},...,m_{q}$ are non-zero integers satisfying $m_{1}+...+m_{q}=M$.
These considerations allow us to rewrite \eqref{eq:polymer expansion phi}
as 
\begin{align}
Z^{{\rm polymer}}= & 1+\sum_{M\geq1}\sum_{q=1}^{M}\frac{1}{q!}\sum_{m_{1}+\cdots+m_{q}=M}\prod_{t=1}^{q}\frac{\phi\left(m_{t}\right)}{m_{t}!}.\nonumber \\
= & 1+\sum_{q=1}^{M}\frac{1}{q!}\sum_{M=q}^{\infty}\sum_{m_{1}+\cdots+m_{q}=M}\prod_{t=1}^{q}\frac{\phi\left(m_{t}\right)}{m_{t}!}\nonumber \\
= & 1+\sum_{q=1}^{M}\frac{1}{q!}\left(\sum_{M=1}^{\infty}\frac{\phi\left(M\right)}{M!}\right)^{q}\nonumber \\
= & \exp\left(\sum_{M=1}^{\infty}\frac{\phi\left(M\right)}{M!}\right).
\end{align}
The logarithm of the polymers partition function $\eqref{eq:Appendix polymer partition function}$
can thus be expressed as

\begin{eqnarray}
\ln Z^{{\rm polymer}} & = & \sum_{M=1}^{\infty}\frac{1}{M!}\sum_{\gamma_{1},...,\gamma_{M}\subset\Gamma}\prod_{k=1}^{M}K\left(\gamma_{k}\right)\nonumber \\
 &  & \times U_{M}\left(\gamma_{1},\cdots,\gamma_{M}\right).
\end{eqnarray}

\section{Activities of Loops and Bounds\label{app:activity bounds}}

\subsection{High-Temperature General Models}

We recall that the partition function of a general factor graph model
is given by \eqref{eq:spinsystem} and the weights by \eqref{eq:hightemperature}.
We take $\beta$ small enough so that 
\begin{align}
\bigr\vert\psi_{a}(s_{\partial a})-1\bigl\vert\leq2\beta\sup_{a\in C}\sum_{I\subset\partial a}\vert J_{I}\vert\equiv\mu,\label{eq:high_temperature}
\end{align}
As will be seen later on we will need $\beta$ small enough so that
$\mu<O(1/l_{\max}^{2}r_{\max})$.

In order to find bounds on the activities \eqref{eq:Ki} and \eqref{eq:Ka},
we should control the behavior of the Belief-Propagation messages.
This is realized through the BP equations \eqref{eq:BP_equation_ai}
and \eqref{eq:BP_equation_ia}. We first choose to parametrize the
BP distributions $\nu,\widehat{\nu}$ with real numbers $\zeta,\widehat{\zeta}$
\begin{equation}
\widehat{\nu}_{ai}(s_{i})=\frac{1+s_{i}\tanh\widehat{\zeta}_{a\rightarrow i}}{2}\text{ and }\nu_{ia}(s_{i})=\frac{1+s_{i}\tanh\zeta_{i\rightarrow a}}{2}.
\end{equation}
The BP equation \eqref{eq:BP_equation_ai} now reads 
\begin{equation}
\tanh\widehat{\zeta}_{a\to i}=\frac{\sum_{s_{\partial a}}\psi_{a}(s_{\partial a})s_{i}\prod_{j\in\partial a\setminus i}\nu_{ja}(s_{j})}{\sum_{s_{\partial a}}\psi_{a}(s_{\partial a})\prod_{j\in\partial a\setminus i}\nu_{ja}(s_{j})}.
\end{equation}
Injecting the high-temperature condition \eqref{eq:high_temperature}
leads to the following bound 
\begin{align}
\vert\tanh\widehat{\zeta}_{a\to i}\vert & \leq\frac{\sum_{s_{\partial a}}\vert\psi_{a}(s_{\partial a})-1\vert\prod_{j\in\partial a\setminus i}\nu_{ja}(s_{j})}{1-\sum_{s_{\partial a}}\vert\psi_{a}(s_{\partial a})-1\vert\prod_{j\in\partial a\setminus i}\nu_{ja}(s_{j})}\nonumber \\
 & \leq\frac{\mu}{1-\mu}\nonumber \\
 & \leq2\mu,\label{eq:bound_zeta_ai}
\end{align}
where in the last line, we use the fact that $\mu<1/2$.

The other BP equation \eqref{eq:BP_equation_ia} takes the form 
\begin{equation}
\tanh\zeta_{i\to a}=\tanh\left(\sum_{b\in\partial i\setminus a}\widehat{\zeta}_{b\to i}\right).
\end{equation}
Using the bound \eqref{eq:bound_zeta_ai} on messages $\widehat{\zeta}_{a\to i}$
gives 
\begin{align}
\vert\tanh\zeta_{i\to a}\vert & \leq\tanh\left((l_{\max}-1)\tanh^{-1}(2\mu)\right)\nonumber \\
 & \leq2(l_{\max}-1)\mu.\label{eq:bound_zeta_ia}
\end{align}
The inequalities \eqref{eq:bound_zeta_ai} and \eqref{eq:bound_zeta_ia}
can be restated in terms of distributions $\widehat{\nu},\nu$ and
take the form 
\begin{equation}
\begin{cases}
\frac{1-2\mu}{2}\leq\widehat{\nu}_{ai}(s_{i})\leq\frac{1+2\mu}{2}\\
\frac{1-2(l_{\max}-1)\mu}{2}\leq\nu_{ia}(s_{i})\leq\frac{1+2(l_{\max}-1)\mu}{2}.
\end{cases}\label{eq:bound_nu}
\end{equation}
By noticing that $\sum_{s}s\widehat{\nu}_{ai}(s)=\tanh\widehat{\zeta}_{a\to i}$
and using the bound \eqref{eq:bound_zeta_ai}, we are in position
to control the activity \eqref{eq:Ka} 
\begin{align}
\vert K_{a}(g)\vert & \leq\frac{\sum_{s_{\partial a}}\vert\psi_{a}(s_{\partial a})-1\vert\prod_{i\in\partial a\setminus g}\nu_{ia}(s_{i})\prod_{i\in\partial a\cap g}\widehat{\nu}_{ai}(-s_{i})}{1-\sum_{s_{\partial a}}\vert\psi_{a}(s_{\partial a})-1\vert\prod_{i\in\partial a}\nu_{ia}(s_{i})}\nonumber \\
 & +\frac{\prod_{i\in\partial a\cap g}\vert\sum_{s_{i}}s_{i}\widehat{\nu}_{ai}(s_{i})\vert}{1-\sum_{s_{\partial a}}\vert\psi_{a}(s_{\partial a})-1\vert\prod_{i\in\partial a}\nu_{ia}(s_{i})}\nonumber \\
 & \leq\frac{\mu+(2\mu)^{\vert\partial a\cap g\vert}}{1-\mu}\nonumber \\
 & \leq6\mu,
\end{align}
where in the last line we use the fact that subgraphs $g$ have no
dangling edges (i.e. $\vert\partial a\cap g\vert\geq2$) and $\mu\leq1/2$.

The second activity \eqref{eq:Ki} is directly controlled using bounds
on distributions $\widehat{\nu}$ and $\nu$ given by equations \eqref{eq:bound_nu}
\begin{align}
\vert K_{i}(g)\vert & \leq\frac{\sum_{s_{i}}\prod_{a\in\partial i\setminus g}\vert\widehat{\nu}_{ai}(s_{i})\vert\prod_{a\in\partial i\cap g}\vert\nu_{ia}(-s_{i})\vert}{\vert\sum_{s_{i}}\prod_{a\in\partial i}\widehat{\nu}_{ai}(s_{i})\vert}\nonumber \\
 & \leq\left(\frac{1+(l_{\max}-1)2\mu}{1-2\mu}\right)^{l_{\max}}\nonumber \\
 & \leq\left(1+4l_{\max}\mu\right)^{l_{\max}}.
\end{align}

The total activity of a generalized loop $g$, given by the relation
\eqref{eq:total_activity}, is then bounded by 
\begin{align}
\vert K(g)\vert & \leq\prod_{i\in g\cap V}\vert K_{i}\vert\prod_{a\in g\cap C}\vert K_{a}\vert\nonumber \\
 & \leq\exp\biggl(\vert g\cap V\vert l_{\max}\ln(1+4l_{\max}\mu)+\vert g\cap C\vert\ln(6\mu)\biggr).\label{eq:high_temp_activity}
\end{align}
There are two antagonistic contributions in the loops activities.
One is exponentially increasing in the number of variable nodes. The
other is exponentially decreasing in the number of check nodes. We
define the size of a subgraph $g$, denoted by $\vert g\vert$, as
the total number of variable and check nodes contained in the loop
\begin{align}
\vert g\vert:=\vert g\cap C\vert+\vert g\cap V\vert.\label{eq:polymer_size}
\end{align}
In order to show that the activities in \eqref{eq:high_temp_activity}
are exponentially decreasing in the loop size, we should show that
the number of variable nodes contained in a loop cannot be arbitrarily
larger than the number of check nodes. Consider the number of edges
contained in a subgraph. We can bound from above this number counted
from the check node perspective and we can find a lower bound counted
from the variable node perspective. This leads to the following bound
on the number of variable nodes 
\begin{align}
r_{\max}\vert g\cap C\vert\geq2\vert g\cap V\vert.\label{eq:variable_check_bound}
\end{align}
Using the definition \eqref{eq:polymer_size} and the bound \eqref{eq:variable_check_bound},
we find that for every non-negative numbers $p\geq0$ and $q\geq0$
\begin{align}
\vert g\cap V\vert p-\vert g\cap C\vert q & =-(p+q)\vert g\cap C\vert+p\vert g\vert\nonumber \\
 & \leq\vert g\vert\frac{r_{\max}p-2q}{2+r_{\max}}
\end{align}
This implies the upper bound for the exponent in \eqref{eq:high_temp_activity}
\begin{align}
\frac{\vert g\vert}{2+r_{{\rm max}}}\ln\biggl((6e\mu)^{2}(1+4l_{{\rm max}}\mu)^{r_{{\rm max}}l_{{\rm max}}}\biggr).\label{eq:eb}
\end{align}
Moreover for $\mu<1/(2l_{\max}^{2}r_{\max})$ we have 
\begin{align}
(6e\mu)^{2}(1+4l_{\max}\mu)^{r_{\max}l_{\max}}\leq(6e\mu)^{2}.\label{eq:fb}
\end{align}
From \eqref{eq:high_temp_activity}, \eqref{eq:eb} and \eqref{eq:fb}
we deduce the bound on the activities 
\begin{align}
\vert K(g)\vert & \leq(6e\mu)^{2\vert g\vert/(2+r_{\max})}.\label{eq:actb}
\end{align}

\subsection{LDGM Codes}

We recall that the partition function of a LDGM code is 
\begin{align}
Z_{{\rm LDGM}} & =\sum_{\underline{s}\in\{-1,1\}^{n}}\prod_{a\in C}e^{h_{a}\prod_{i\in\partial a}s_{i}}.
\end{align}
The LDGM codes can be seen as a special case of the high-temperature
general models with $I\to\partial a$ and $\beta J_{I}\to h_{a}$.
The high-temperature condition translates into $2\sup_{a}\vert h_{a}\vert=\mu\ll1$.
Recalling that $\vert h_{a}\vert=h=\frac{1}{2}\ln\frac{1-p}{p}$,
we see that the high-temperature condition is equivalent to taking
$p$ close to $1/2$. The bound on the activity is obtained by applying
\eqref{eq:actb}

\begin{align}
\vert K(g)\vert & \leq\left(6e\ln\frac{1-p}{p}\right)^{2\vert g\vert/(2+r_{\max})}.\label{eq:ldgm_activity_bound}
\end{align}
The activities of the LDGM codes have a high-temperature bound and
the high-noise regime $p\approx1/2$ is then similar to a high-temperature
regime for general models.

There is a remarkable simplification for LDGM ensembles with no degree
one check nodes. In this case the BP equations admit a trivial fixed
point where $\nu_{ia}(s)=1/2$, $\widehat{\nu}_{ia}(s)=1/2$. The
activities at the trivial fixed point can be computed exactly 
\begin{align}
K_{a}^{\text{trivial}}(g)=\begin{cases}
\tanh h_{a} & \hspace{0.5cm}\text{if}\hspace{0.25cm}\partial a\cap g=\partial a\\
0 & \hspace{0.5cm}\text{otherwise},
\end{cases}
\end{align}
and 
\begin{align}
K_{i}^{\text{trivial}}(g)=\begin{cases}
1 & \hspace{0.3cm}\text{if \ensuremath{\vert\partial a\cap g\vert}is even}\\
0 & \hspace{0.3cm}\text{otherwise}.
\end{cases}
\end{align}
Subgraphs contributing in the loop sum are only those which have check
nodes with maximal induced degree and variable nodes with odd degree.
Their activities admit the simple bound 
\begin{align}
\vert K(g)^{\text{trivial}}\vert\leq(1-2p)^{\vert g\cap C\vert}\leq(1-2p)^{\frac{2\vert g\vert}{2+r_{\max}}}
\end{align}

\subsection{Regular LDPC Codes}

The LDPC codes cannot be seen as high-temperature models. Their partition
function 
\begin{equation}
Z_{{\rm LDPC}}=\sum_{\underline{x}\in\{0,1\}^{n}}\prod_{a\in C}\mathbb{I}\left(\oplus_{i\in\partial a}x_{i}=0\right)\prod_{i\in V}\exp((-1)^{x_{i}}h_{i}).
\end{equation}
is composed of two type of weights. The variable node weights, coming
from channel observations, satisfies a high-temperature condition
at high noise. But the check node weights, enforcing a parity check
constraint, always admit a configuration of variable which cancels
the weight. Thus it make impossible that the check node weights satisfies
the high-temperature condition \eqref{eq:high_temperature}.

We use the standard parametrization for the BP distributions $\nu,\widehat{\nu}$
in term of the real numbers $\eta,\widehat{\eta}$

\begin{align}
\widehat{\nu}_{ai}(s_{i})=\frac{1+s_{i}\tanh\widehat{\eta}_{a\rightarrow i}}{2}\text{ and }\nu_{ia}(s_{i})=\frac{1+s_{i}\tanh\eta_{i\rightarrow a}}{2}.
\end{align}
With this parametrization the Belief-Propagation equations \eqref{eq:BP_equation_ai},
\eqref{eq:BP_equation_ia} for the messages reads 
\begin{equation}
\begin{cases}
\tanh(\widehat{\eta}_{a\rightarrow i})=\prod_{j\in\partial a\setminus i}\tanh\eta_{j\rightarrow a}\\
\eta_{i\rightarrow a}=h_{i}+\sum_{b\in\partial i\setminus a}\widehat{\eta}_{b\rightarrow i}.
\end{cases}\label{eq:LDPC_BP_equations}
\end{equation}
Indeed the BP equations always admit the trivial solution $\tanh\eta_{a\to i}=1$,
$\tanh\widehat{\eta}_{i\to a}=1$. Thus unlike the high-temperature
cases, the BP equations of LDPC codes are not sufficient to control
the BP fixed points. We need an extra requirement on the class of
fixed point used in the loop expansion, called high-noise fixed points.

Given $\epsilon>0$, we say that a fixed point $(\eta,\widehat{\eta})$
is a $\epsilon$ high-noise fixed point if for all $(i,a)\in E$ 
\begin{align}
\vert\tanh\eta_{i\rightarrow a}\vert\leq\theta.\label{eq:high_noise_condition}
\end{align}
where 
\begin{equation}
\theta=(1+\epsilon)\tanh h.
\end{equation}
The condition \eqref{eq:high_noise_condition} can be justified by
looking at the Taylor expansion of solution at high noise. For $h=0$,
the BP equations \eqref{eq:LDPC_BP_equations} admit the simple solution
$\tanh\eta_{a\to i}=0$, $\tanh\widehat{\eta}_{i\to a}=0$. If we
compute the Taylor expansion of this solution with respect to the
noise parameter, we find 
\begin{equation}
\begin{cases}
\tanh\widehat{\eta}_{a\rightarrow i}=\prod_{j\in\partial a\setminus i}\tanh h_{j}\\
\tanh\eta_{i\rightarrow a}=\tanh h_{i}+\sum_{b\in\partial i\setminus a}\prod_{j\in\partial a\setminus i}\tanh h_{j},
\end{cases}
\end{equation}
plus some term of order $\mathcal{O}((\tanh h)^{r})$. This shows
that there exists a $h_{0}(\epsilon,n)$ such that the high-noise
condition \eqref{eq:high_noise_condition} is satisfied for $h<h_{0}(\epsilon,n)$.
However it does not guaranteed that $h_{0}(\epsilon,n)$ is uniform
in the size of the graph.

By using the high-noise condition \eqref{eq:high_noise_condition}
along with the BP equations \eqref{eq:LDPC_BP_equations}, we find
the reciprocal bound on messages from check nodes to variable nodes
\begin{align}
\vert\tanh\widehat{\eta}_{a\to i}\vert\leq\theta^{r-1}.\label{eq:high_noise_condition_reciprocal}
\end{align}

We recall that the induced degree of check and variable node are denoted
by $d_{a}(g)=\vert\partial a\cap g\vert$ and $d_{i}(g)=\vert\partial i\cap g\vert$
respectively. The number of check nodes and variable nodes with prescribed
induced degree by $n_{s}\left(g\right)=\left\vert \left\{ i\in g\cap V\vert d_{i}(g)=s\right\} \right\vert $
and $m_{t}\left(g\right)=\left\vert \left\{ a\in g\cap C\vert d_{a}(g)=t\right\} \right\vert $.
For LDPC codes, the activities associated with check nodes \eqref{eq:Ka}
are 
\begin{align}
K_{a}(g)=\frac{u_{a}+(-1)^{d_{a}(g)}v_{a}}{1+u_{a}w_{a}},
\end{align}
where 
\begin{align}
\begin{cases}
u_{a} & =\prod_{i\in\partial a\setminus g}\tanh\eta_{i\rightarrow a}\\
v_{a} & =\prod_{i\in\partial a\cap g}\tanh\widehat{\eta}_{a\rightarrow i}\\
w_{a} & =\prod_{i\in\partial a\cap g}\tanh\eta_{i\rightarrow a}.
\end{cases}
\end{align}
Using inequalities \eqref{eq:high_noise_condition}, \eqref{eq:high_noise_condition_reciprocal},
it is straightforward to bound the check activities 
\begin{align}
\vert K_{a}(g)\vert & \leq\frac{\vert u_{a}\vert+\vert v_{a}\vert}{1-\vert u_{a}\vert\vert w_{a}\vert}\nonumber \\
 & \leq\frac{\theta^{r-d_{a}(g)}+\theta^{(r-1)d_{a}(g)}}{1-\theta^{r}}.
\end{align}
Thus for a fixed numerical constant $\alpha_{1}$ that we can take
close to one 
\begin{equation}
\vert K_{a}(g)\vert\leq\left\{ \begin{array}{lcl}
1+\alpha_{1}\theta^{r} & \mbox{if} & d_{a}(g)=r\\
\alpha_{1}\theta^{r-d_{a}(g)} & \mbox{if} & d_{a}(g)\neq r.
\end{array}\right.
\end{equation}

The activities associated with variable nodes \eqref{eq:Ki} reads
\begin{align}
K_{i}(g)=\frac{e^{(u_{i}-v_{i})}+(-1)^{d_{i}(g)}e^{-(u_{i}-v_{i})}}{e^{(u_{i}+w_{i})}+e^{-(u_{i}+w_{i})}}\prod_{a\in\partial i\cap g}\frac{\cosh\widehat{\eta}_{a\to i}}{\cosh\eta_{i\to a}},
\end{align}
where 
\begin{align}
\begin{cases}
u_{i} & =h_{i}+\sum_{a\in\partial i\setminus g}\widehat{\eta}_{a\rightarrow i}\\
v_{i} & =\sum_{a\in\partial i\cap g}\eta_{i\rightarrow a}\\
w_{i} & =\sum_{a\in\partial i\cap g}\widehat{\eta}_{a\rightarrow i}.
\end{cases}
\end{align}
Again by a direct application of inequalities \eqref{eq:high_noise_condition}
and \eqref{eq:high_noise_condition_reciprocal}, we find 
\begin{align}
\vert K_{i}(g)\vert & \leq\frac{e^{(d_{i}(g)+1)\tanh^{-1}\theta}+(-1)^{d_{i}(g)}e^{-(d_{i}(g)+1)\tanh^{-1}\theta}}{e^{(d_{i}(g)+1)\tanh^{-1}\theta}+e^{-(d_{i}(g)+1)\tanh^{-1}\theta}}\nonumber \\
 & \times\left(\frac{1+\theta^{2}}{1-\theta^{2}}\right)^{d_{i}(g)/2}.
\end{align}
For a fixed constant $\alpha_{2}$ close to one we have the following
bound 
\begin{align}
\vert K_{i}(g)\vert\leq\left\{ \begin{array}{lcl}
1+\frac{\alpha_{2}}{2}(1+4d_{i}(g)+d_{i}(g)^{2})\theta^{2} & \mbox{if} & d_{i}(g)\text{ even}\\
\alpha_{2}(1+d_{i}(g))\theta & \mbox{if} & d_{i}(g)\text{ odd}.
\end{array}\right.
\end{align}

Using the formulas \eqref{eq:total_activity} we derive the following
estimate of subgraphs activities for $\theta<\theta_{0}$ small enough
\begin{equation}
\vert K(g)\vert\leq\overline{K}(\underline{n}(g),\underline{m}(g)),\label{eq:actibound}
\end{equation}
where 
\begin{align}
\overline{K}(\underline{n}(g),\underline{m}(g)) & =\left(1+\alpha_{1}\theta^{r}\right)^{m_{r}(g)}\prod_{t=2}^{r-1}\left(\alpha_{1}\theta^{r-t}\right)^{m_{t}(g)}\nonumber \\
\times & \prod_{\substack{s=2,\\
\mathrm{even}
}
}^{l-1}\left(1+\frac{\alpha_{2}}{2}(1+4s+s^{2})\theta^{2}\right)^{n_{s}(g)}\nonumber \\
\times & \prod_{\substack{s=3,\\
\mathrm{odd}
}
}^{l}\left(\alpha_{2}(1+s)\theta\right)^{n_{s}(g)}.\label{eq: bound activity}
\end{align}

Estimate (\ref{eq: bound activity}) is essentially optimal for small
$\theta$ as can be checked by Taylor expanding $K(g)$ in powers
of $\theta$.

\section{On the Number of Rooted Polymer\label{app:On-the-number}}

The convergence criterion in the polymer expansion requires an evaluation
of the ``entropy'' of rooted polymers. The term ``entropy'' has
to be understood as the number of polymers of a given size. The following
lemma gives a bound on the entropy of polymers on a $d$-regular graph.
Its generalization to irregular bipartite graphs with degrees of variable
nodes and check nodes bounded by $l_{max}$ and $r_{max}$ is straightforward
by setting $d=\max\left(l_{{\rm max}},r_{{\rm max}}\right)$.
\begin{lem}[Bound on the number of rooted polymers]
\label{lem:AP Bound on the number of rooted polymers}Let $\Gamma=\left(V,E\right)$
be a $d$-regular graph with vertex set $V$ and edge set $E$. The
number of polymers $\gamma$ (connected subgraphs) of size $\left|\gamma\cap V\right|=t$
rooted to any vertex $x\in V$ is upper-bounded by
\[
\sum_{\gamma\ni x}\mathbb{I}\left(\left|\gamma\cap V\right|=t\right)\leq e^{dt}.
\]
\end{lem}
\begin{IEEEproof}
A polymer $\gamma\ni x$ is uniquely determined by one of its spanning
tree $T_{\gamma}$ plus the complementary set of edges $\gamma\setminus T_{\gamma}$.
Figure \eqref{fig:AP polymers spanning tree} shows an example of
this injective mapping.
\begin{figure}[tbh]
\centering{}\includegraphics[scale=0.3]{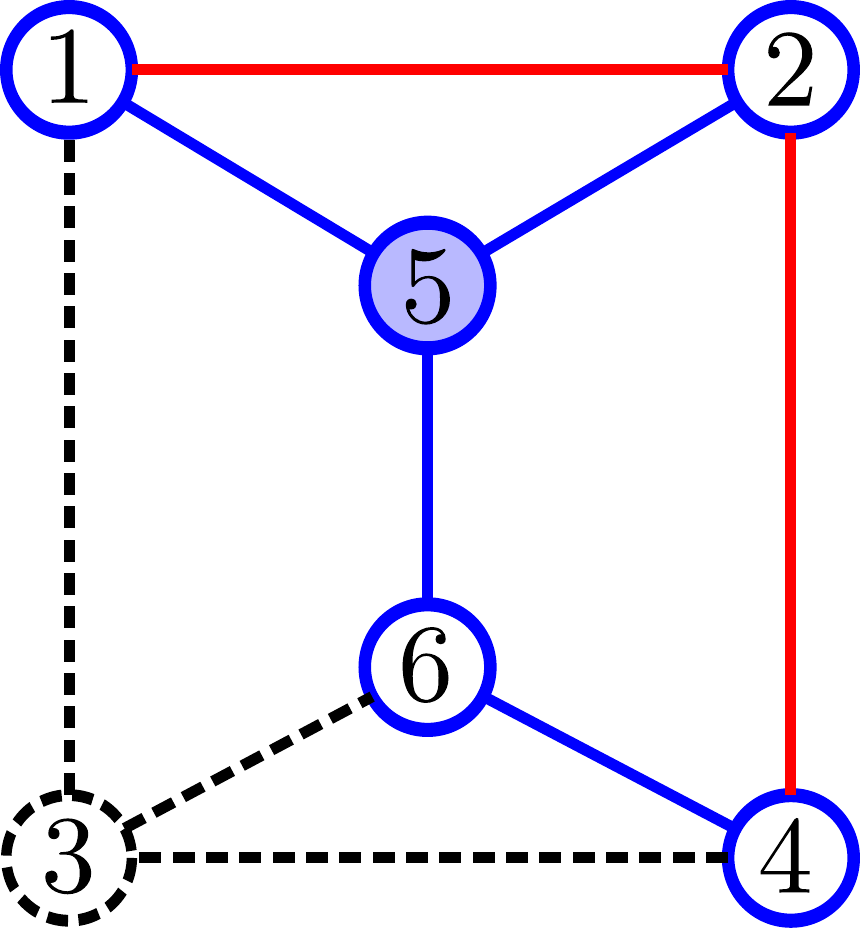}\hfill{}\includegraphics[scale=0.3]{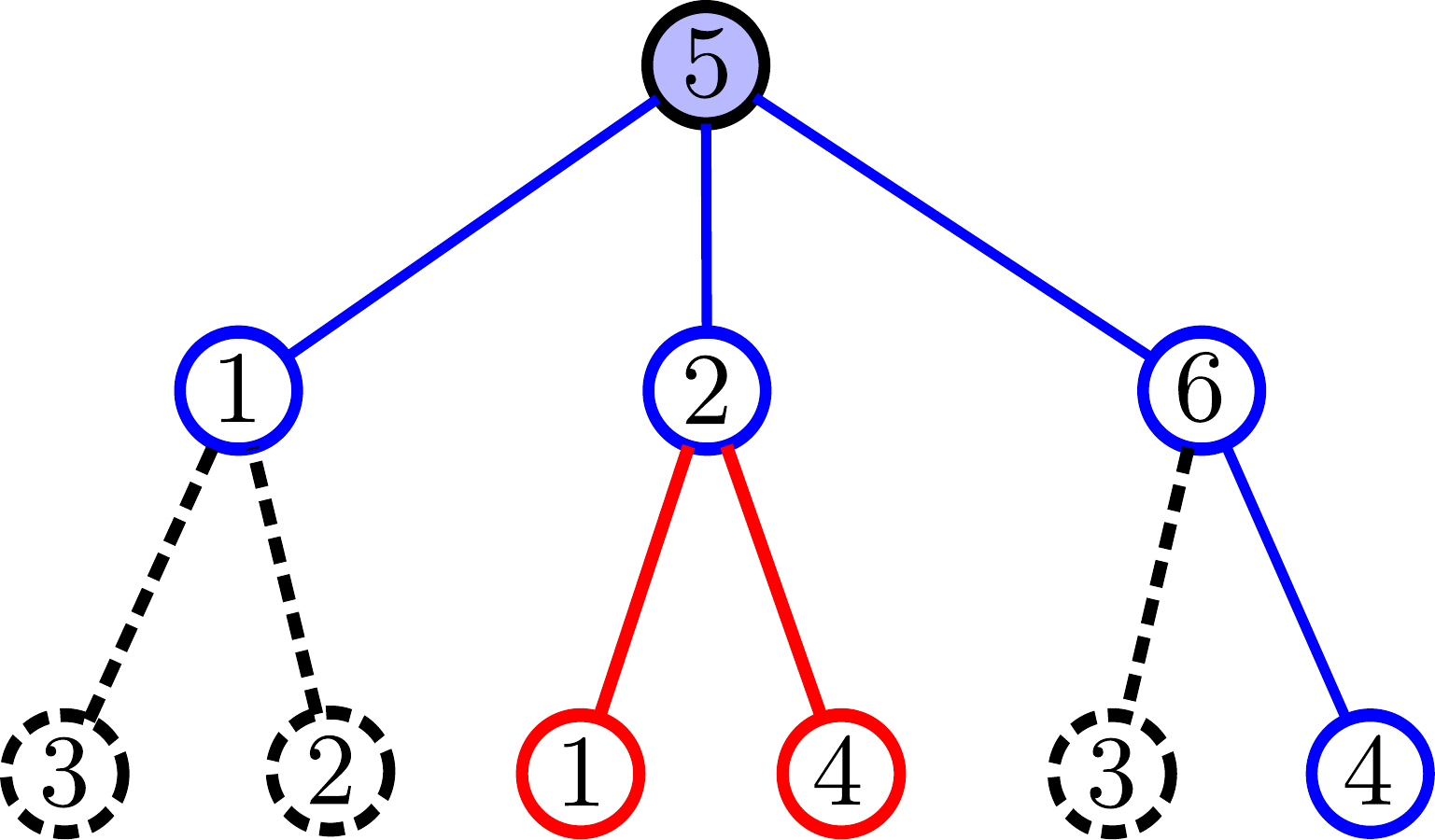}\protect\caption{\selectlanguage{american}%
\label{fig:AP polymers spanning tree}\foreignlanguage{english}{On
the left: a polymer is represented with colored solid lines. A spanning
tree is shown in blue and the complementary edges in red. On the right:
the spanning tree is shown on the computational tree in blue with
a possible representation of the complementary edges in red. }\selectlanguage{english}%
}
\end{figure}
We ask the following question: If $T$ is a spanning tree, how many
different polymers have $T$ as a spanning tree? In other words how
many combinations of complementary edges can be made once $T$ is
given? Let $g$ be the graph spanned by $T$ which contains the most
edges. As $T$ is a tree, $\left|T\cap E\right|=\left|T\cap V\right|-1$.
Therefore the number of complementary edges unspecified by a spanning
tree is at most
\begin{align}
\left|g\setminus T\right| & =\left|g\cap E\right|-\left|T\cap E\right|\nonumber \\
 & =\left|g\cap E\right|-\left|T\cap V\right|+1\nonumber \\
 & =\left|g\cap E\right|-\left|g\cap V\right|+1\nonumber \\
 & \leq\left(\frac{d}{2}-1\right)\left|g\cap V\right|+1.\label{eq:AP number complentary edges}
\end{align}
Denote by $A_{t}\left(x\right)$ the number of polymers of size $t$
rooted in $x\in V$ and call $B_{t}$ the number of rooted $d$-ary
trees with size $t$. Based on the previous considerations
\begin{equation}
A_{t}\left(x\right)\leq2^{\left(\frac{d}{2}-1\right)t+1}B_{t}.\label{eq:AP polymer to spanning tree}
\end{equation}
To find a formula for $B_{t}$, we use a derivation based on generating
functions similar to the one of Catalan numbers in \cite{wilf2006generatingfunctionology}.
Define the generating function
\begin{equation}
B\left(z\right):=\sum_{t=0}z^{t}B_{t}.\label{eq:AP tree generating function}
\end{equation}
If one removes the root of a $d$-ary tree it splits the tree into
$d$ trees of smaller size. This yields the following equation for
the generating function 
\begin{equation}
B=1+zB^{d}.\label{eq:AP recurence generating fonction}
\end{equation}
By using the Lagrange-Bürmann formula on Equation \eqref{eq:AP recurence generating fonction}
we find
\begin{equation}
B_{t}=\frac{1}{t\left(d-1\right)+1}\binom{td}{t}.
\end{equation}
Finally we can relax the bound \eqref{eq:AP polymer to spanning tree}
to have a simpler expression by noticing that
\begin{align}
2^{\left(\frac{d}{2}-1\right)t+1}B_{t}\leq & e^{dt}.
\end{align}

\end{IEEEproof}

\section{Proof of Lemma \ref{th:McK lemma}\label{app:MCKay}}

In this appendix we prove the Lemma \ref{th:McK lemma} which is restated
below for convenience \begin{lemma*} Fix $\delta>0$. Assume $l\geq3$
odd and $l<r$. There exists a constant $C>0$ depending only on $l$
and $r$ such that for $\theta$ small enough 
\begin{equation}
\mathbb{P}_{\Gamma}\biggl[\vert R_{{\rm large}}\vert\geq\delta\biggr]\leq\frac{1}{\delta}e^{-Cn},
\end{equation}
where 
\begin{equation}
R_{{\rm large}}=\sum_{g\subset\Gamma{~{\rm s.t~}}\exists\gamma\subset g{~{\rm with~}}\vert\gamma\vert\geq\lambda n}K(g).
\end{equation}
\end{lemma*} 
\begin{IEEEproof}
Let $\Omega_{\Gamma}\left(\underline{n},\underline{m}\right)$ be
the set of all $g\subset\Gamma$ with prescribed type $(\underline{n}(g),\underline{m}(g))$.
By (\ref{eq: bound activity}) and the Markov bound 
\begin{align}
\mathbb{P}\left[\vert R_{{\rm large}}\vert\geq\delta\right]\leq\frac{1}{\delta}\sum_{\underline{n},\underline{m}\in\Delta}\overline{K}\left(\underline{n},\underline{m}\right)\mathbb{E}_{\Gamma}\left[\left\vert \Omega_{\Gamma}\left(\underline{n},\underline{m}\right)\right\vert \right],\label{eq:activity_vs_entropy}
\end{align}
where 
\begin{align}
\Delta\equiv\biggl\{\left(\underline{n},\underline{m}\right)\mid & \lambda n\leq\sum_{s=2}^{l}n_{s}+\sum_{t=2}^{r}m_{t},\sum_{s=2}^{l}sn_{s}=\sum_{t=2}^{r}tm_{t},\nonumber \\
 & \sum_{s=2}^{l}n_{s}<n,\sum_{t=2}^{r}m_{t}<nl/r\biggr\}.\label{eq:delta_nm}
\end{align}
The expectation of the number of $g\subset\Gamma$ with prescribed
type can be estimated by combinatorial bounds provided by McKay \cite{mckay2010subgraphs}.
It turns out that these subgraphs proliferate exponentially in $n$
only for a subdomain of $\Delta$ where $\overline{K}\left(\underline{n},\underline{m}\right)$
is exponentially much smaller in $n$. In the subdomain where $\overline{K}\left(\underline{n},\underline{m}\right)$
is not small (but it is always bounded) the number of subgraphs is
sub-exponential when $l$ is odd and $l<r$. As a consequence for
$l$ odd and $l<r$, we are able to prove that the sum on the right
hand side of \eqref{eq:activity_vs_entropy} is smaller than $e^{-Cn}$.

Let us now give the details of this calculation. Let $\omega=4r^{2}-2r+2$,
a number independent of $n$. The combinatorial bound is only valid
for subgraphs $g$ with number of edges at most equal to $nl-\omega$.
Thus we have to separate the domain of summation \eqref{eq:delta_nm}
into 
\begin{equation}
\Delta_{\omega}=\Delta\cap\left\{ \left(\underline{n},\underline{m}\right)\mid\sum_{s=2}^{l}sn_{s}\leq nl-\omega\right\} \,\text{and}\,\Delta_{\omega}^{c}=\Delta\setminus\Delta_{\omega},
\end{equation}
and handle each part separately.

For $\left(\underline{n},\underline{m}\right)\in\Delta_{\omega}^{c}$
a trivial bound of the expected size of $\Omega_{\Gamma}\left(\underline{n},\underline{m}\right)$
is given by 
\begin{align}
\mathbb{E}_{\Gamma}\left[\left\vert \Omega_{\Gamma}\left(\underline{n},\underline{m}\right)\right\vert \right] & \leq\binom{nl}{\omega}\nonumber \\
 & =O\left(n^{\omega}\right).
\end{align}
This is nothing but simply counting the possible subgraphs obtained
by removing $\omega$ edges from $\Gamma$. For the same reason the
activity \eqref{eq: bound activity} is upper-bounded by 
\begin{align}
\left|\overline{K}\left(\underline{n},\underline{m}\right)\right| & \leq\left(1+\alpha_{1}\theta^{r}\right)^{n\frac{l}{r}}\left(1+\frac{\alpha_{2}}{2}(1+4l+l^{2})\theta^{2}\right)^{\omega}\nonumber \\
 & \times\left(\alpha_{2}(1+l)\theta\right)^{n-\omega}\nonumber \\
 & =O\left(\left(\alpha_{2}\left(1+l\right)\left(1+\alpha_{1}\right)^{\frac{l}{r}}\theta\right)^{n-\omega}\right).\label{eq:K_delta_c_omega}
\end{align}
Indeed the activity of the total graph is upper-bounded by $\overline{K}\left(\Gamma\right)=\left(1+\alpha_{1}\theta^{r}\right)^{n\frac{l}{r}}\left(\alpha_{2}(1+l)\theta\right)^{n}$.
The worst case scenario for the activity of subgraphs obtained by
removing $\omega$ edges from $\Gamma$ is then bounded by \eqref{eq:K_delta_c_omega}.
Therefore for every $\theta<\theta_{1}=\left(\alpha_{2}\left(1+l\right)\left(1+\alpha_{1}\right)^{\frac{l}{r}}\right)^{-1}$
and $n$ large enough, there exists a constant $C_{1}>0$ depending
on $l$ and $r$ such that 
\begin{equation}
\sum_{\underline{n},\underline{m}\in\Delta_{\omega}^{c}}\overline{K}\left(\underline{n},\underline{m}\right)\mathbb{E}_{\Gamma}\left[\left\vert \Omega_{\Gamma}\left(\underline{n},\underline{m}\right)\right\vert \right]\leq e^{C_{1}\left(n-\omega\right)\ln\left(\frac{\theta}{\theta_{1}}\right)}.\label{eq:sum_K_Omega_delta_c}
\end{equation}

For $\left(\underline{n},\underline{m}\right)\in\Delta_{\omega}$,
the probability that a graph $g$ with prescribed type $(\underline{n}(g),\underline{m}(g))$
belongs to $\Omega_{\Gamma}(\underline{n},\underline{m})$ is upper-bounded
by McKay's estimate 
\begin{align}
\mathbb{P}_{\Gamma}\left[g\in\Omega_{\Gamma}(\underline{n},\underline{m})\right] & \leq\frac{\prod_{s=2}^{l}\left(\frac{l!}{\left(l-s\right)!}\right)^{n_{s}}\prod_{t=2}^{r}\left(\frac{r!}{\left(r-t\right)!}\right)^{m_{t}}}{\frac{(nl-\omega)!}{\left(nl-\sum_{s=2}^{l}sn_{s}-\omega\right)!}}.
\end{align}
By counting the number of graph $g$ with prescribed degrees $(\underline{n}(g),\underline{m}(g))$,
we deduce 
\begin{align}
\mathbb{E}_{\Gamma}\left[\left\vert \Omega_{\Gamma}(\underline{n},\underline{m})\right\vert \right] & \leq\binom{n}{n_{2},...,n_{l}}\binom{n\frac{l}{r}}{m_{2},...,m_{r}}\nonumber \\
 & \times\frac{\left(\sum_{s=2}^{l}sn_{s}\right)!}{\prod_{s=2}^{l}s!^{n_{s}}\prod_{t=2}^{r}t!^{m_{t}}}\mathbb{P}_{\Gamma}\left[g\in\Omega_{\Gamma}(\underline{n},\underline{m})\right].\label{eq:bound_on_number_of_loop}
\end{align}
Setting $x_{s}=\frac{n_{s}}{n}$, $y_{t}=\frac{r}{l}\frac{m_{t}}{n}$,
we perform an asymptotic analysis for $n$ large of the bound \eqref{eq:bound_on_number_of_loop}.
Therefore we transform factorials using Stirling approximation valid
for $k>0$ 
\begin{equation}
e^{\frac{1}{12k+1}}\leq\frac{k!}{\sqrt{2\pi k}e^{-k}k^{k}}\leq e^{\frac{1}{12k}}.\label{eq:Stirling bound}
\end{equation}
In order to simplify the terms in $\omega$ we also use the following
inequality valid for $n>l\omega$ and $0\leq z\leq1-\frac{\omega}{nl}$
\begin{align}
\left(1-z\right)\ln\left(1-z\right)-\frac{\omega}{nl}\ln\frac{\omega}{nl} & \geq\left(1-z-\frac{\omega}{nl}\right)\ln\left(1-z-\frac{\omega}{nl}\right).\label{eq:entropy omega bound}
\end{align}
This could be easily proven by considering a joint probability distribution
$p(A=0,B=0)=0$, $p(A=0,B=1)=z$, $p(A=1,B=0)=\frac{\omega}{nl}$,
$p(A=1,B=1)=1-z-\frac{\omega}{nl}$ and applying the inequality 
\[
H\left(A\right)\leq H\left(A,B\right),
\]
where $H$ is the Shannon entropy in nat.

Observe that
\begin{equation}
-\left(1-\frac{\omega}{nl}\right)\ln\left(1-\frac{\omega}{nl}\right)\leq\frac{\omega}{nl}.\label{eq:entropy vs ln bound}
\end{equation}
Using the relations \eqref{eq:bound_on_number_of_loop}, \eqref{eq:Stirling bound},
\eqref{eq:entropy omega bound} along with \eqref{eq:entropy vs ln bound}
gives the following bound on the number of subgraphs of $\Gamma$
\begin{equation}
\mathbb{E}_{\Gamma}\left[\left\vert \Omega_{\Gamma}(\underline{n},\underline{m})\right\vert \right]\leq C_{l,r}n^{\frac{\omega}{l}+2}\exp\left(nlf\left(\underline{x},\underline{y}\right)\right),\label{eq:mckay_exp_form}
\end{equation}
where $C_{l,r}$ is a constant that depends only on $l$ and $r$
and 
\begin{align}
f\left(\underline{x},\underline{y}\right) & =\left(1-\sum_{s=2}^{l}\frac{s}{l}x_{s}\right)\ln\left(1-\sum_{s=2}^{l}\frac{s}{l}x_{s}\right)\nonumber \\
 & +\left(\sum_{s=2}^{l}\frac{s}{l}x_{s}\right)\ln\left(\sum_{s=2}^{l}\frac{s}{l}x_{s}\right)\nonumber \\
 & +\frac{1}{l}\left(\sum_{s=2}^{l}x_{s}\ln\binom{l}{s}\right)+\frac{1}{r}\left(\sum_{t=2}^{r}y_{t}\ln\binom{r}{t}\right)\nonumber \\
 & -\frac{1}{r}\left(\left(1-\sum_{t=2}^{r}y_{t}\right)\ln\left(1-\sum_{t=2}^{r}y_{t}\right)+\sum_{t=2}^{r}y_{t}\ln y_{t}\right)\nonumber \\
 & -\frac{1}{l}\left(\left(1-\sum_{s=2}^{l}x_{s}\right)\ln\left(1-\sum_{s=2}^{l}x_{s}\right)+\sum_{s=2}^{l}x_{s}\ln x_{s}\right).
\end{align}
The bound on the activity \eqref{eq: bound activity} can also be
put in a form where the growth rate in $n$ is explicit 
\begin{equation}
\overline{K}\left(\underline{n},\underline{m}\right)=\exp\left(nlk_{\theta}\left(\underline{x},\underline{y}\right)\right),\label{eq:activity_exp_form}
\end{equation}
where 
\begin{align}
k_{\theta}\left(\underline{x},\underline{y}\right) & =\frac{y_{r}}{r}\ln\left(1+\alpha_{1}\theta^{r}\right)+\sum_{t=2}^{r-1}\frac{y_{t}}{r}\ln\left(\alpha_{1}\theta^{r-t}\right)\nonumber \\
 & +\sum_{_{\substack{s=2,\\
\mathrm{even}
}
}}^{l-1}\frac{x_{s}}{l}\ln\left(1+\frac{\alpha_{2}}{2}\left(1+4s+s^{2}\right)\theta^{2}\right)\nonumber \\
 & +\sum_{_{_{\substack{s=3,\\
\mathrm{odd}
}
}}}^{l}\frac{x_{s}}{l}\ln\left(\alpha_{2}\left(1+s\right)\theta\right).
\end{align}
Define the ensemble 
\begin{align}
\Delta' & \equiv\left\{ \left(\underline{x},\underline{y}\right)\in\mathbb{R}_{+}^{l-1}\times\mathbb{R}_{+}^{r-1}\mid\lambda\leq\frac{1}{l}\sum_{s=2}^{l}x_{s}+\frac{1}{r}\sum_{t=2}^{r}y_{t},\right.\nonumber \\
 & \left.\sum_{s=2}^{l}\frac{s}{l}x_{s}=\sum_{t=2}^{r}\frac{t}{r}y_{t},\sum_{s=2}^{l}x_{s}<1,\sum_{t=2}^{r}y_{t}<1\right\} .\label{eq:delta_prime}
\end{align}
It is easy to verify that if $\left(\underline{n},\underline{m}\right)\in\Delta_{\omega}$
then $\left(\underline{x},\underline{y}\right)\in\Delta'$. Combining
\eqref{eq:sum_K_Omega_delta_c}, \eqref{eq:mckay_exp_form} and \eqref{eq:activity_exp_form}
gives finally 
\begin{equation}
\sum_{\underline{n},\underline{m}\in\Delta_{\omega}}\overline{K}\left(\underline{n},\underline{m}\right)\mathbb{E}_{\Gamma}\left[\left\vert \Omega_{\Gamma}\left(\underline{n},\underline{m}\right)\right\vert \right]\leq C'_{l,r}n^{\frac{\omega}{l}+l+r}\exp\left(nl\Lambda\right),\label{eq:sum_K_Omega_delta}
\end{equation}
where 
\begin{equation}
\Lambda\left(\theta\right)=\max_{\left(\underline{x},\underline{y}\right)\in\Delta'}\left\{ f\left(\underline{x},\underline{y}\right)+k_{\theta}\left(\underline{x},\underline{y}\right)\right\} .\label{eq:Lambda_max}
\end{equation}
In \eqref{eq:sum_K_Omega_delta_c} we estimate the sum over $\left(\underline{n},\underline{m}\right)\in\Delta_{\omega}$
by the crude bound $\left|\Delta_{\omega}\right|\leq n^{l-1}\left(\frac{nl}{r}\right)$$^{r-1}$.

It remains now to prove that $\Lambda\left(\theta\right)$ is strictly
negative for $\theta$ small enough. In the subspace $\Delta_{0}\subset\Delta'$
defined by having all coordinates $x_{s}$ for $s$ odd and $y_{t}$
for $t<r$ equal to zero, the function $k_{\theta}\left(\underline{x},\underline{y}\right)$
can be made arbitrarily close to zero as $\theta$ is small. Notice
also that in the complementary subspace $\Delta'\setminus\Delta_{0}$,
the function $k_{\theta}\left(\underline{x},\underline{y}\right)$can
be made arbitrarily negative for small $\theta$ due to the presence
of the terms $\ln\theta$. It is therefore sufficient to show that
the restriction of $f\left(\underline{x},\underline{y}\right)$ to
$\Delta_{0}$ is strictly negative. Call $z_{s}=x_{2s}$ and define
the set 
\begin{equation}
\Delta'_{0}\equiv\left\{ \underline{z}\in\mathbb{R}_{+}^{\frac{l-1}{2}}\mid l\lambda\leq\sum_{s=1}^{\frac{l-1}{2}}z_{s}<1\right\} .
\end{equation}
If $\underline{z}\in\Delta'_{0}$ then $\left(\underline{x},\underline{y}\right)\in\Delta_{0}$,
as we can express the variable $y_{r}=\sum_{s=1}^{\frac{l-1}{2}}\frac{2s}{l}z_{s}$
with the second constraint in \eqref{eq:delta_prime}. The restriction
to $\Delta_{0}$ of $f\left(\underline{x},\underline{y}\right)$ can
be recast into the form 
\begin{equation}
lf\left(\underline{x},\underline{y}\right)=f_{0}\left(\underline{z}\right)-\left(1-\frac{l}{r}\right)h_{2}\left(\sum_{s=1}^{\frac{1-1}{2}}\frac{2s}{l}z_{s}\right),
\end{equation}
where 
\begin{align}
f_{0}\left(\underline{z}\right) & =-\left(l-1\right)h_{2}\left(\sum_{s=1}^{\frac{l-1}{2}}\frac{2s}{l}z_{s}\right)\nonumber \\
 & +\left(\sum_{s=1}^{\frac{l-1}{2}}z_{s}\ln\binom{l}{2s}\right)\nonumber \\
 & -\left(\left(1-\sum_{s=1}^{\frac{l-1}{2}}z_{s}\right)\ln\left(1-\sum_{s=1}^{\frac{l-1}{2}}z_{s}\right)+\sum_{s=1}^{\frac{l-1}{2}}z_{s}\ln z_{s}\right).
\end{align}
The function $f_{0}$ takes its maximum in $\Delta'_{0}$ at $\underline{z}^{*}=\frac{1}{2^{l-1}}\binom{l}{2s}$
and $f_{0}$ $\left(\underline{z}^{*}\right)=0.$ Thus, since $2\lambda<\sum_{s=1}^{\frac{l-1}{2}}\frac{2s}{l}z_{s}<1-\frac{1}{l}$,
for $\left(\underline{x},\underline{y}\right)\in\Delta_{0}$ we have
\begin{equation}
lf\left(\underline{x},\underline{y}\right)<-\left(1-\frac{l}{r}\right)\min\left\{ h_{2}\left(2\lambda\right),h_{2}\left(\frac{1}{l}\right)\right\} <0.\label{eq:max_fxy}
\end{equation}
Therefore for $\theta$ small enough $\Lambda\left(\theta\right)<0$
and there exist for large $n$ a constant $C_{2}>0$ depending on
$l$ and $r$ such that 
\begin{equation}
\sum_{\underline{n},\underline{m}\in\Delta_{\omega}}\overline{K}\left(\underline{n},\underline{m}\right)\mathbb{E}_{\Gamma}\left[\left\vert \Omega_{\Gamma}\left(\underline{n},\underline{m}\right)\right\vert \right]\leq e^{-nC_{2}}.\label{eq:sum_K_Omega_delta_final}
\end{equation}
Combining Markov's inequality \eqref{eq:activity_vs_entropy} and
inequalities \eqref{eq:sum_K_Omega_delta_c}, \eqref{eq:sum_K_Omega_delta_final}
ends the proof.

Notice that the condition $\frac{l}{r}<1$ appears naturally in \eqref{eq:max_fxy}.
It is thus necessary that the graph $\Gamma$ describes a code (i.e.
with positive rate).
\end{IEEEproof}

\noindent{\bf Acknowledgements.} This work was performed while M. Vuffray was at EPFL. He acknowledges support from 
the Swiss national Foundation grant no 140388. The authors thank M. Chertkov, R. Urbanke and P. Vontobel for discussions 
and comments. 

\bibliographystyle{ieeetr}
\bibliography{refloop}

\end{document}